\documentclass[twocolumn,showpacs,prb,superscriptaddress,aps,floatfix]{revtex4-1}

\usepackage{graphicx}% Include figure files
\usepackage{dcolumn}% Align table columns on decimal point
\usepackage{bm}% bold math
\usepackage{nicefrac}
\usepackage{hyperref}% add hypertext capabilities
\usepackage{gensymb}  %to get deg symbol for example
\usepackage{amsmath}
\usepackage{xcolor}
\usepackage{amssymb}
\usepackage{soul}
\usepackage{float}
\usepackage{multirow}

\def\kk         {{\bf k}}
\def\qq		{{\bf q}}

\newcommand{\modena}{Department of Physics, Computer Science, and Mathematics, University of Modena and Reggio Emilia, Modena, 41125, Italy}
\newcommand{\piim}{Aix Marseille Univ, CNRS, PIIM, Physique des Interactions Ioniques et Moléculaires, UMR 7345, Marseille, France}
\newcommand{\etsf}{European Theoretical Spectroscopy Facility (ETSF)}

\newcommand{\cdymer}{\(\mathrm{C}_{\mathrm{B}}\mathrm{C}_{\mathrm{N}}\)}
\newcommand{\CsubB}{\(\mathrm{C}_{\mathrm{B}}\)}
\newcommand{\CsubN}{\(\mathrm{C}_{\mathrm{N}}\)}
\newcommand{\vacb}{\(\mathrm{V}_{\mathrm{B}}\)}
\newcommand{\vacn}{\(\mathrm{V}_{\mathrm{N}}\)}
\newcommand{\BsubN}{\(\mathrm{B}_{\mathrm{N}}\)}
\newcommand{\NsubB}{\(\mathrm{N}_{\mathrm{B}}\)}
\newcommand{\wbn}{$w$BN}
\newcommand{\hbn}{$h$BN}
\newcommand{\cbn}{$c$BN}

\begin{document}

\title{Wurtzite Boron Nitride as a potential defects host}% Force line breaks with \\
\author{Martino Silvetti}
\email{martino.silvetti@univ-amu.fr}
\affiliation{\piim}
\affiliation{\modena}
\affiliation{\etsf}
\author{Matilde Pouyot}
\affiliation{\piim}
%\author{Fulvio Paleari}%
%\affiliation{\cnrmodena}
%\affiliation{\etsf}
%\author{Adam Gali}
%\affiliation{\wigner}
%\affiliation{\bute}
\author{Elena Cannuccia}
\affiliation{\piim}
\affiliation{\etsf}
\date{\today}% It is always \today, today,%  but any date may be explicitly specified

\begin{abstract}
Wurtzite boron nitride (\wbn) is a polymorph of boron nitride and serves as an intermediate phase in the transition from \hbn\, to \cbn\, under high pressure and temperature conditions. Owing to these extreme synthesis conditions, \wbn\, likely inherits defects from hexagonal phase, where bright and stable single-photon emitters have been observed in both the visible and ultraviolet spectral ranges. 
	%such as vacancies, antisites, and carbon impurities, which strongly influence its optical and electronic properties. 
While \hbn\, and cubic BN (cBN) have been extensively studied for hosting quantum emitters, \wbn\, remains comparatively unexplored. In this work, we use first-principles hybrid DFT calculations to investigate the formation energies and electronic structures of key native point defects in \wbn, including boron and nitrogen vacancies, antisites, and carbon impurities, across various charge states. Our results reveal the potential of \wbn\, as a robust platform for optically active defects. These properties make it a promising candidate for quantum technologies operating under extreme conditions. This study lays the groundwork for future experimental efforts in defect identification and engineering in \wbn.
\end{abstract}

%\keywords{Suggested keywords}%Use showkeys class option if keyword display desired
\maketitle

%\tableofcontents

%======================================
\section{\label{sec:intro}Introduction}

%The native defects create discrete energy levels in the energy gap of the semiconductor. These could be either shallow (acceptors or donors) or deep (traps) energy levels.

Wurtzite Boron Nitride (\wbn), one of the boron nitride polymorphs, was synthesized for the first time in 1963 by Bundy and Wentorf\,\cite{bundy_wentorf_wbn_first_synt} 
and it plays the role of intermediary in the phase transition between the hexagonal (\hbn) and the cubic (\cbn) phase of boron nitride under high pressure, where the hexagonal 
to wurtzite phase transition is completed at around 13 GPa\,\cite{segura}. For a long time it was considered to be a metastable material that could not exist
at ambient pressure\,\cite{solozhenko,Tani_1975}.
However, more recently, the successful synthesis of stable, millimeter-sized samples of \wbn\ has been reported\,\cite{Deura_2017,segura}. 
Chen et al.,\cite{Chen2019} proposed that planar faults may play a major role in the stabilization process.
%Furthermore, it has been suggested by Gunning et al. that stabilisation can be achieved by alloying\,\cite{wbn_alloying,tsao2018ultrawide}.

Due to the complex high-pressure and high-temperature growth techniques, there is a high probability that imperfections originally present in \hbn\, are retained in \wbn. 
In addition to dislocations and stacking faults, the most common atomic-scale imperfections are vacancies, antisites and interstitials. These atomic imperfections, known as native defects, are thermodynamically stable and can exist at any temperature above absolute zero. Their concentration depends on the formation energy and the temperature: lower formation energies and higher temperatures lead to higher defect concentrations. However, the concentration of a given native defect species also depends on the specific semiconductor material. 

Because these defects are frozen in the crystal at the growth temperature, the defect type with the highest concentration strongly influences the material's properties.  
Furthermore, native defects can exist in different charge states, which depend on the specific atomic site they occupy within the crystal lattice. 
%Due to the charge state of the native defects in III–V compounds, high-purity crystals grown without any intentional impurities (dopants) exhibit electron or hole concentrations more than their intrinsic charge carrier densities. The electrical conductivity type (p-type or n-type) is decided by the dominant native defect configuration in the material.

The literature on experimental investigations of the electronic and optical properties of $w$BN counts still few papers. In Ref. ~\cite{yixi}, authors have reported 
the experimental dielectric constant, and affirm that the optical gap was found at $8.7\pm0.5$. Some theoretical and computational studies have been carried
out using density functional theory (DFT)\,\cite{ks-equations} and the quasi-particle
formalism in the \(G_{0}W_{0}\) approximation\,\cite{strinati,hedin,onida_reining_rubio}, where it is established that \wbn\,is a wide band gap material with a direct and indirect band gap at \(G_{0}W_{0}\) level of 10.02 eV and 6.4 eV, respectively\,\cite{silvetti23,christensen,cappellini}. 

A theoretical study of \wbn\, as a potential host of native point defects is still lacking, despite similar investigations have been conducted for \hbn\,\cite{Weston2018,orellana2001stability} and \cbn\,\cite{orellana2001stability,tararan_zobelli}. In \hbn\, the characterization of the chemical nature of the native point defects and their thermodynamics properties was triggered by the 2016 discovery that \hbn\, can host bright and stable single-photon emitters in the visible\cite{Martinez2016} and ultraviolet spectral ranges\cite{Bourrellier2016}. 
From the energetic point of view isolated vacancies and antisites have been shown to possess very high formation energies and are therefore unlikely to 
form in large concentrations, whereas carbon impurities have relatively low formation energies and may be present in significant concentration in 
undoped \hbn\,\cite{Weston2018}. 
The experimental evidence of the presence of carbon during the growth process is the ubiquitous $4.08$ eV emission line in \hbn. The carbon dimer has been proposed\,\cite{mackoit2019} as responsible for this emission line accompanied by a unitary quantum efficiency and a very short radiative lifetime, leading to a potential candidate as quantum emitter in the near UV. Present with three charge states, ($q=-1, 0$ and $1$), the neutral one is the most stable one, with a formation energy of 2.2 eV throughout the band gap and there are two defect states in the band gap.

Similarly, cubic boron nitride (cBN), another wide-bandgap and superhard boron nitride polymorph, has been extensively investigated as a host for optically and magnetically active point defects\cite{orellana2001stability,nguyen2024computationally}. In particular, cBN has demonstrated promising features such as deep defect levels, high thermal stability, and mechanical resilience, which are advantageous for quantum applications under extreme conditions. These studies underscore the importance of exploring other boron nitride allotropes—such as wBN—that share similar structural and electronic characteristics, yet remain comparatively less understood.

In this work, we employ first-principles methods based on hybrid density functional theory to provide information on the energetic and electronic properties of the most relevant native point defects in wurtzite boron nitride (\wbn). Our comprehensive study, based on the supercell method, focuses on boron and nitrogen vacancies, boron and nitrogen antisites, carbon substitutions, and the carbon dimer. We have considered different charge states and included proper finite cell size corrections. These investigations may pave the way for experimental characterization of intrinsic defects hosted in \wbn\, crystals in order to further optimize the properties of this material in the context of quantum technological applications, following similar efforts previously conducted for \hbn\, and \cbn. In particular, the exceptional hardness of \wbn, which is comparable to that of \cbn, makes it a promising candidate for hosting defect levels suitable for quantum technologies operating under extreme conditions, such as high pressure or mechanical stress\,\cite{albe,orellana}.

This manuscript is structured as follows: in section~\ref{sec:comput_det} we briefly review the computational methods; section~\ref{sec:formen_ctl} we present the
results on the formation energies and the charge transition levels. In Sec.\,\ref{sec:results} for each defect we present structural and electronic properties. 
Finally in Sec.\,\ref{concl} we give conclusions and perspectives.

%=======================================
\section{Computational Setup}
\label{sec:comput_det}
%======================================
In the following we provide information on the computational setup that we adopted first focusing on the unit cell and then on the defective supercell.\\
\\
\emph{Pure unit cell}
%\marginpar{\textcolor{red} {Put citations and double check the details}}

\begin{figure*}
	\centering
	\includegraphics[scale=0.45]{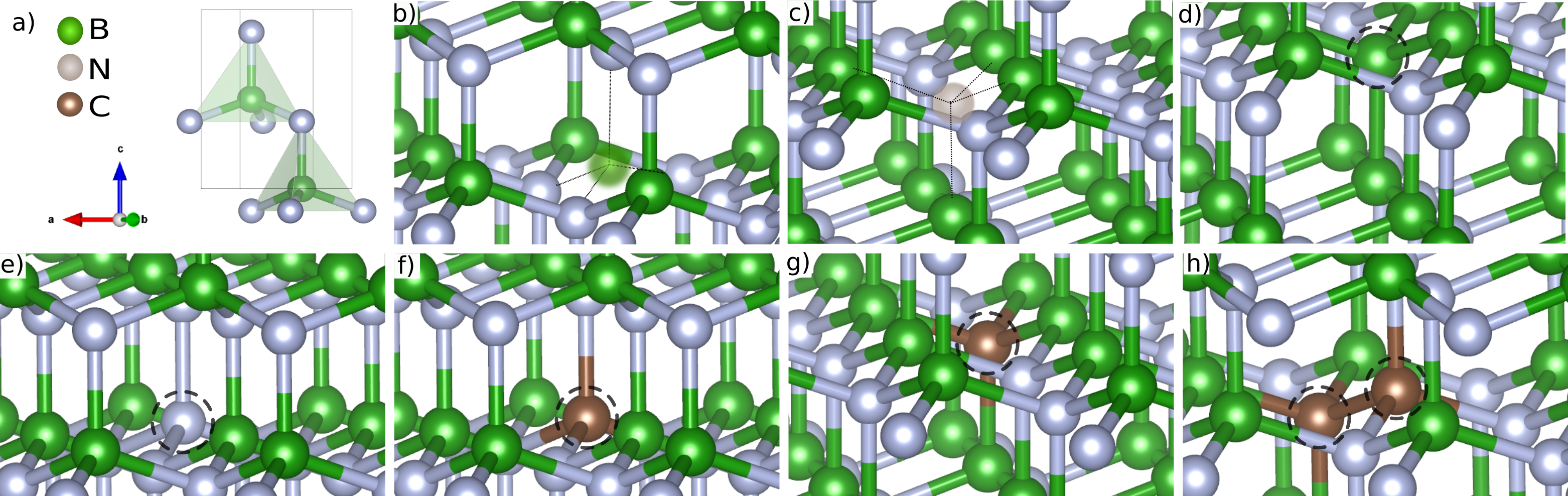}
	\caption{Structural configuration of all defects considered. a) Pure $w$BN unit cell, b) and c) Boron and Nitrogen vacancies, d) and e) $\mathrm{B}_\mathrm{N}$ and $\mathrm{N}_\mathrm{B}$ antisites, f) and g) \CsubB\, and \CsubN\, substitutionals, h) \(\mathrm{C}_{\mathrm{B}}\mathrm{C}_{\mathrm{N}}\) carbon dimer.}
	\label{fig:AllDef_Struct}
\end{figure*}

We performed preliminary density functional calculations (DFT) on the $w$BN unit cell, (shown in Fig.~\ref{fig:AllDef_Struct}(a)), using a plane-wave basis set, as implemented in the Quantum ESPRESSO package\cite{Giannozzi_2017}. A plane wave kinetic energy cutoff of 70 Ry was applied and the Brillouin zone was sampled with a 12$\times$12$\times$8 $\kk$-grid. This gave total energies that converged within 4 meV per valence electron. The optimized atomic positions were determined under the convergence criterion for forces set to $10^{-5}$ Ry/Bohr. First, we used the PBE exchange-correlation functional\cite{perdew1996generalized} with optimized norm-conserving Vanderbilt pseudopotentials\cite{VANSETTEN201839} and obtained direct and indirect band gap values of 8.35 eV and 5.27 eV, respectively, in agreement with previously published results\,\cite{silvetti23}. Semi local functionals like PBE underestimate the gap. In order to provide an accurate band structure for the host in this work  
we used a hybrid functional based on the Gaussian Attenuation Scheme (GAU)\,\cite{song2013communication} which incorporates short-range Hartree-Fock (HF) exchange with PBE exchange-correlation functional. In the following we will refer to it as GAU-PBE. The main advantage of GAU-PBE functional is the absence of divergence at $\qq=0$, making it well-suited for systems requiring accurate treatment of localized electronic states. 

When using hybrid functionals, convergence of the unit cell total energy and the electronic gap with respect to the \qq-grids is required. The convergence is considered reached with a 12$\times$12$\times$8 $\kk$-grid and a 9$\times$9$\times$6 \qq-grid. The direct and indirect band gap values are 10.16 eV and 6.81 eV respectively. The direct gap occurs along the $\Gamma$-M line at the $X=(0,1/12,0)$ point, while the indirect gap occurs between points $\Gamma$ and $K$. \\
To assess the quality of the band structure obtained with GAU-PBE, we compared it with the results obtained using the GW method\cite{aryasetiawan1998gw}.
The $G_0W_0$ calculation were performed using 50 bands for the expansion of G and W, a cutoff of 7800 mRy for the dielectric constant and the Godby-Needs plasmon-pole model\,\cite{PPA} for the dynamical part of W.   
In Tab.~\ref{tab:Egap_Hyb_vs_GGA} we report the direct and indirect band gap obtained with the two approaches and functional. We found a difference of 21~meV for the direct band gap and 34~meV for the indirect one that we consider acceptable, considering also the fact that $G_0W_0$ usually underestimates the gap in large band-gap materials\,\cite{Gant2022} due to the overscreening at DFT level.

%%%%%%%%%%%%%%%%%%%%%%%%%%%%%%%%%%%%%%%%%%%%%%%%%%%%%%TABLE 1 %%%%%%%%%%%%%%%%%%%%%%%%%%%%%%%%%%%%%%%%%%%%%%%%%%%%%%%%%%%%%%%%%%%%%%%%%%%%%%%%%%%%%%
\begin{table}[h]
        \centering
	\begin{tabular}{c|c|c}
		\hline
		    & $E_{dir}$ (eV) & $E_{ind}$ (eV) \\
		\hline
		PBE     & 8.35  & 5.27  \\
		GAU-PBE  & 10.16  & 6.81 \\
		$G_0W_0$@PBE   & 9.95  & 6.47 \\
		\hline
	\end{tabular}
	\caption{Direct and indirect energy band gaps with standard PBE functionals, GAU-PBE hybrid functionals and $G_0W_0$ approximation.} 
	\label{tab:Egap_Hyb_vs_GGA}
\end{table}

%%%%%%%%%%%%%%%%%%%%%%%%%%%%%%%%%%%%%%%%%%%%%%%%%%%%%%FIGURE 1 %%%%%%%%%%%%%%%%%%%%%%%%%%%%%%%%%%%%%%%%%%%%%%%%%%%%%%%%%%%%%%%%%%%%%%%%%%%%%%%%%%%%%%

%%%%%%%%%%%%%%%%%%%%%%%%%%%%%%%%%%%%%%%%%%%%%%%%%%%%%%%%%%%%%%%%%%%%%%%%%%%%%%%%%%%%%%%%%%%%%%%%%%%%%%%%%%%%%%%%%%%%%%%%%%%%%%%%%%%%%%%%%%%%%%%%%%%%%
\emph{From the pure to the defective supercell} 

 %Our aim is to determine the position of defect levels within the band gap of $w$BN.
Starting from a relaxed \wbn\, unit cell containing four atoms, we constructed a pristine supercell of size N$_x\times$N$_y\times$N$_z$, which allowed us to fold the K point of the Brillouin zone, where the conduction bands minimum occurs, onto the $\Gamma$ point. Having both the valence band maximum and the conduction band minimum at $\Gamma$ point is essential for accurately determining the position of defect levels within the band gap of \wbn, which is the primary objective of this study. We then chose a $6\times6\times4$ supercell containing 576 atoms. We used 70 Ry as kinetic energy cut-off and sampled the Brillouin zone at the Gamma point in our DFT calculations. 

The defective supercell was created by introducing a defect (adding or removing atoms from the supercell), shown in Fig.~\ref{fig:AllDef_Struct} and labeled in the following $X^q$, where the superscript $q$ is the charge state. Spin polarization was taken into account for all the results presented in this work.

We followed several standard steps for each charge state of a given defect, and variations are described where necessary. First, we relaxed the internal coordinates by imposing an initial guess on the magnetization of each atom, with a Gaussian smearing of $10^{-4}$~Ry. At this stage, the geometry and total magnetization are optimized simultaneously. For certain charge states, we realized that Gaussian smearing equal to $10^{-3}$~Ry was needed to achieve convergence, albeit at the cost of fractional occupancies. In these cases, suspecting Jahn-Teller distortion of the geometry around the defect site, we searched for a more stable configuration by manually breaking the defect symmetry and searching for a new energy minimum at PBE level. Once the relaxed geometries were found, we proceeded with a standard DFT calculation using either PBE or GAU-PBE hybrid functionals.

%and we converged the \(\boldsymbol{k}\)-grid sampling to have the occupancies as close as possible either to 0 or 1. Then we carry out a new relaxation of the geometry by imposing the total magnetization exited in the previous step and without applying a smearing. This second calculation is carried out only for the point \(\Gamma\). The third stage consist of a non self-consistent calculation to find the defect wavefunctions.  For each defect charge state studied we report the convergence results for the aforementioned stages in the supplemental materials. 

%======================================
\section{Formation Energy and Charge Transition Levels}
\label{sec:formen_ctl}
%======================================
Once the relaxed geometries had been obtained, a self-consistent field (SCF) calculation was performed to determine the energy levels of the defects within the perfect lattice band gap. In this context, we are assuming that the size of the supercell used is such that the average potentials at the boundaries of the defective and the pristine system are numerically close, i.e. referenced to a similar value for the vacuum. This calculation also determined the total magnetization, total energy, and level occupations. Stability of the defect \(X^q\) is assessed by looking at the formation energy\cite{rev_freysoldt_defects, zhang_northrup_formation_en} \(E_{f}[X^{q}](E_{F})\) as a function of the Fermi energy \(E_{F}\) given by the expression:
\begin{multline}
	E_{f}[X^{q}](E_{F})=E^{def}_{tot}[X^{q}]-E^{bulk}_{tot}+E^{q}_{dil}+\\-\sum_{i}n_{i}\mu_{i}+q(E_{F}+E_{VB}),
	\label{eq:formation_energy}
\end{multline}
\begin{figure*}
	\centering
	\includegraphics[scale=0.3]{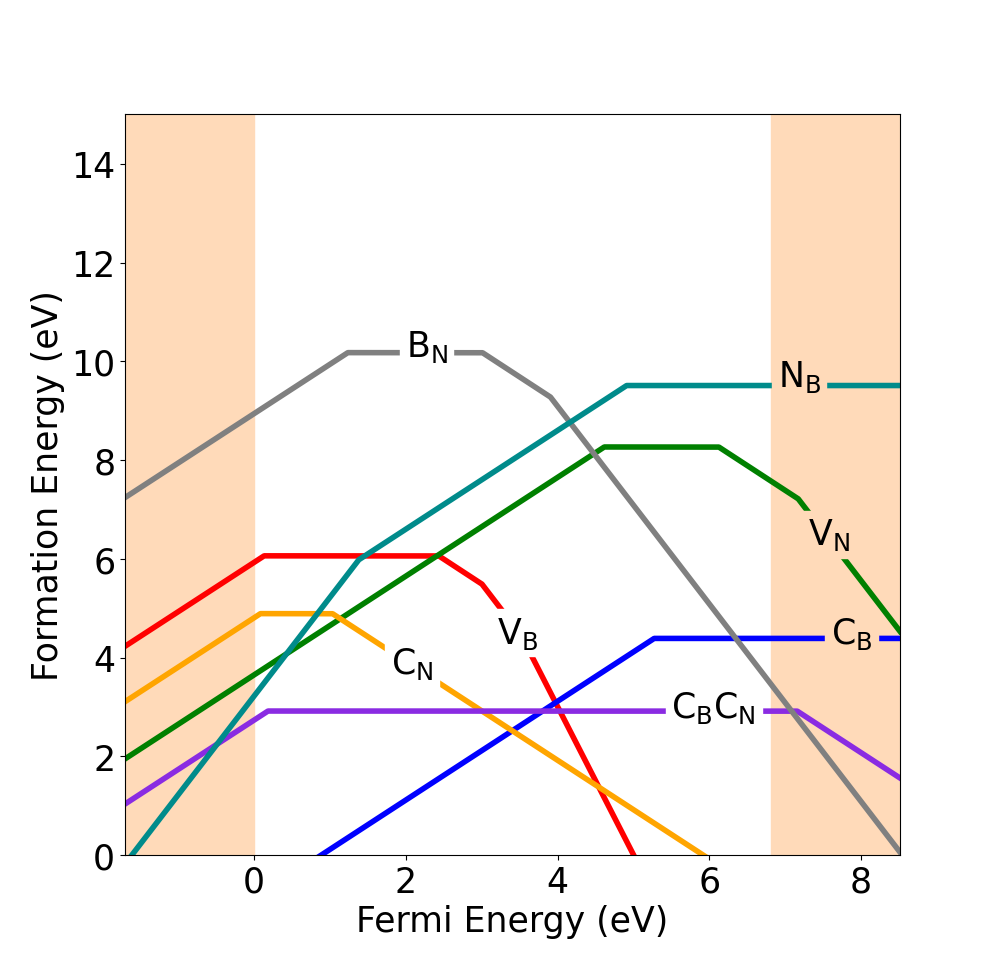}
	\includegraphics[scale=0.3]{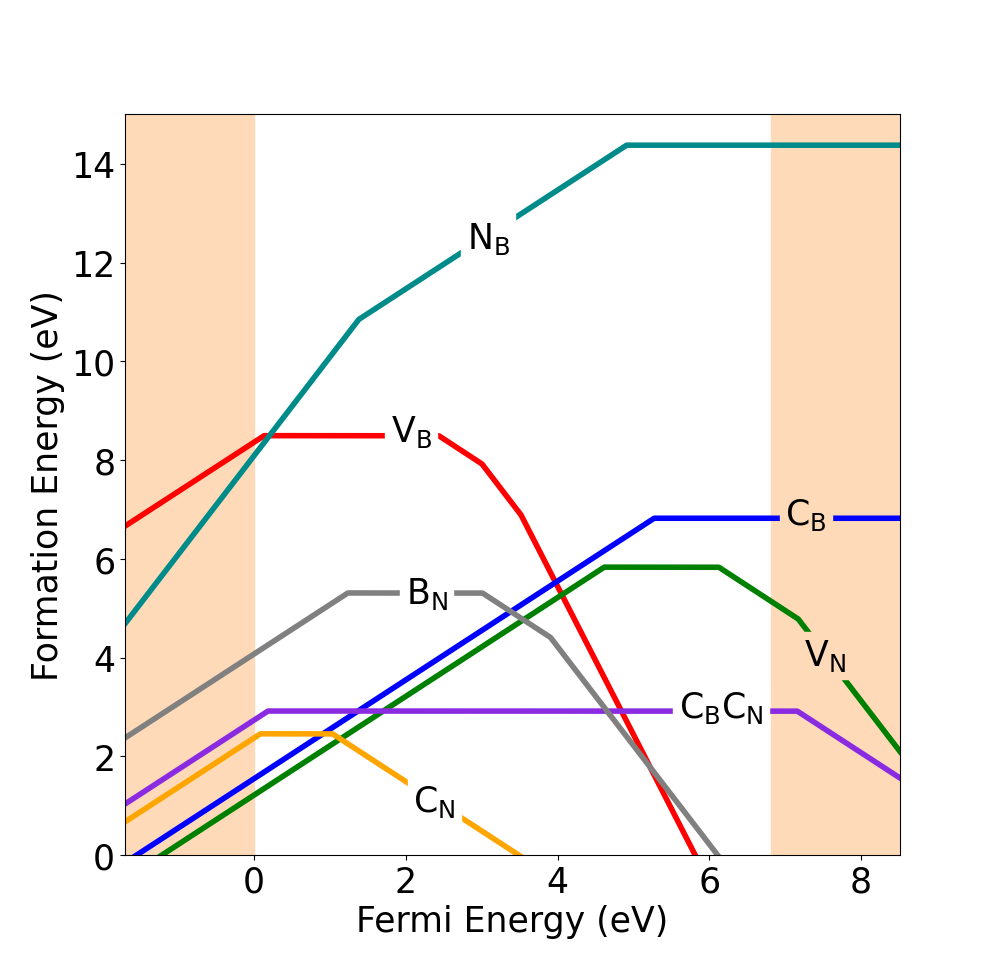}
	\caption{\label{fig:formation_energies_GAU_NB-rich_limits} Formation energies at the GAU-PBE level as function of the Fermi energy \(E_{F}\) for seven defects in the N-rich limit (on the left) and in the B-rich limit (on the right). The slope of each line corresponds to the charge state (see Eq.\ref{eq:formation_energy}).}
\end{figure*}
where \(E_{tot}^{def}[X^{q}]\) and \(E_{tot}^{bulk}\) are the total energies of the defective and pure supercells respectively, \(n_{i}\) is the number of atoms of the species \(i\) added (\(n_{i}>0\)) or removed (\(n_{i}<0\)) and exchanged with a reservoir of chemical potential \(\mu_{i}^{ref}\), for the elemental species $i$, in order to form the defect. Then \(\mu_{i}\) is the chemical potential of the atomic species $i$, it is a variable which represents the experimental growth conditions and it cannot be higher than $\mu_i^{ref}$. The Fermi energy \(E_{F}\) is the chemical potential of the electrons therefore \(q(E_{F}+E_{VB})\) is the energy needed to charge the defect referenced to the valence band maximum \(E_{VB}\). 
In order to account the spurious interactions between the periodic images of the charged defects, the corrective term $E^{q}_{dil}$ is introduced. We address this point by following the Freysoldt-Neugebauer-Van de Walle (FNV) scheme \cite{freysoldt_phys_rev_lett,Freysoldt_phys_stat_b}as implemented in the \texttt{sxdefectalign} code\cite{sxdefectalign}.
In Tabs.~\ref{tab_corrections} we report the corrections obtained at PBE and GAU-PBE level for all the defect studied in this work. We observe that at GAU-PBE level the correction term in almost all charge states is larger than that obtained at PBE level. As a general trend we observe that the energy correction doesn't increase linearly with the charge state $q$ and that the correction is positive\,\cite{kumagai_oba_defects}.    
%\EC{Che possiamo scrivere qui a proposito della positivita della correzione? della non simmetria tra + 1 e -1 per un dato difetto? per la dipendenza in $q^2$?} 
\begin{table}
	\begin{tabular}{c|c|c|c}
		defect & Charge state & \(E^{q}_{dil}\) (eV) (PBE) & \(E^{q}_{dil}\) (eV) (GAU-PBE)\\
		\hline
		\multirow{4}{1em}{\(\mathrm{V}_{\mathrm{B}}\)} & +1 & 0.096& 0.097\\
		                                               & -1 & 0.231 & 0.240\\
		                                               & -2 & 0.838 & 0.837\\
		                                               & -3 & 1.837  & 1.868 \\
        \hline
        \multirow{4}{1em}{\(\mathrm{V}_{\mathrm{N}}\)} & +1 & 0.162 & 0.165 \\
                                                       & -1 & 0.217 & 0.226 \\
                                                       & -2 & 0.779 & 0.860\\
        \hline
		\multirow{2}{1em}{\(\mathrm{B}_{\mathrm{N}}\)} & +1  & 0.172 & 0.184 \\
		                                             & -1  & 0.212 & 0.236 \\
							     & -2  & 0.812 & 0.801 \\
	\hline
		\multirow{2}{1em}{\(\mathrm{N}_{\mathrm{B}}\)} & +2 & 0.780 & 0.788\\
		                                                    & +1 & 0.189 & 0.192\\
	%							    & -1 & 0.280 & \\
	\hline
		\multirow{2}{1em}{\(\mathrm{C}_{\mathrm{B}}\)} & +1 & 0.199  & 0.202 \\
                                                       & -1 & 0.080 & 0.076 \\
        \hline
		\multirow{2}{1em}{\(\mathrm{C}_{\mathrm{N}}\)} & +1 & 0.124 & 0.152 \\
                                                       & -1 & 0.202 & 0.205 \\
        \hline
		\multirow{2}{3em}{\(\mathrm{C}_{\mathrm{B}}\mathrm{C}_{\mathrm{N}}\)} & +1 & 0.143 & 0.153 \\
                                                                              & -1 & 0.131 & 0.132 \\
        \hline

	\end{tabular}
	\caption{Charge correction terms for native defects in dilute limit via the FNV scheme, calculated at PBE and GAU-PBE levels. Corrections for charge state \(q=0\) are always zero. \label{tab_corrections}}
\end{table} 
\\

\emph{Reference Chemical Potentials}

The chemical potential is defined as the total energy of species $i$ per number of atoms, $\mu_i=E_{tot}/N_i$. 

In our work, $\mu_\mathrm{B}$ is referenced to the total-energy per atom of Boron in its solid-phase. The solid state structure of pure boron poses problems since it can appear in many polymorphs. Boron often presents itself as an admixture of several phases, the most stable being the \(\alpha\)-rhombohedral\,\cite{pure_boron_alpha_vs_beta} which belongs to the hexagonal system and it is composed of 12 atoms. 
%We converged the total energy value, $E_{tot}$, against the cut-off energy \(E_{cut}\) and the \(\boldsymbol{k}\)-grid. For \(\alpha\)-rhombohedral boron we found that the convergence at GAU-PBE level is reached with a 4\(\times\)4\(\times\)4 \(\boldsymbol{q}\)-grid, and 4\(\times\)4\(\times\)4 \(\boldsymbol{k}\)-grid. The total energy obtained is consistent with previous calculations (see for example Ref.~\cite{Li1992_alphaB}).
$\mu_\mathrm{N}$ is referenced to the total energy of an N atom in the N$_{2}$ molecule. 
%In this case we relaxed the atomic positions to get the optimized interatomic distance, whereas for solid state bulk elements we relaxed also the volume.
In the case of \cdymer, \CsubN\, and \CsubB, $\mu_\mathrm{C}$ is referenced to the total energy of a C atom in diamond. 

After having optimized the atomic positions, we converged the total energy with respect to the \(E_{cut}\) and the \(\boldsymbol{\mathrm{k,q}}\)-grid, as already explained for the unit cell. In table \ref{sumup_chem_pot} we report the referential chemical potentials for each species at PBE and GAU-PBE levels together with their convergence parameters. 
\begin{table}
	\begin{tabular}{|c|c|c|c|c|c|c|c|}
		\hline
		Elemental &\(E_{cut} \) & \(\boldsymbol{\mathrm{k}}\)-grid & \(\boldsymbol{\mathrm{q}}\)-grid & $\mu_\mathrm{i}^{ref}$ (eV)  & $\mu_\mathrm{i}^{ref}$ (eV)  \\
		 Phase    &   (Ry)              &       &                         &  PBE       &  GAU-PBE    \\
		\hline
		\(\alpha\)-B & 60 & 4\(\times\)4\(\times\)4 & 4\(\times\)4\(\times\)4 &  -77.17 &  -77.28\\
		N\({}_{2}\)  & 80 & 2\(\times\)2\(\times\)2 & 1\(\times\)1\(\times\)1 &  -270.81 &  -271.26 \\
		diamond      & 70 & 7\(\times\)7\(\times\)7 & 5\(\times\)5\(\times\)5 &  -155.01 &  -155.29 \\
		\hline
	\end{tabular}
	\caption{Calculated reference chemical potentials of the species $i=\mathrm{B,N,C}$, at PBE and GAU-PBE levels, for each elemental phase with the corresponding cutoff kinetic energy values and $\mathbf{k}$ / $\mathbf{q}$ grids. \label{sumup_chem_pot}}
\end{table}

We define $\Delta \mu_\mathrm{B}$ and $\Delta \mu_\mathrm{N}$ with respect to the reference chemical potentials. The formation enthalpy for \wbn\, is 
\begin{equation}
	\Delta H_f({w\mathrm{BN}})= \Delta \mu_\mathrm{B}+\Delta \mu_\mathrm{N}  
\end{equation}
or equivalently
\begin{equation}
	\Delta H_f(w\mathrm{BN})= \mu_{\mathrm{wBN}}-\mu_\mathrm{B}^{ref}-\mu_\mathrm{N}^{ref}
\end{equation}
as $ \mu_{w\mathrm{BN}}=\mu_\mathrm{B}+\mu_\mathrm{N}$. We obtain $\Delta H_f({w\mathrm{BN}})=-2.34$ eV and $-2.43$  eV at PBE and GAU-PBE levels respectively.
\\

In the following we give a brief explanation of how the growth conditions are taken into account in the calculation of the formation energies.\cite{Abdi2018}
In nitrogen rich conditions, the nitrogen atoms in \wbn\, are assumed to be in equilibrium with $\mathrm{N}_2$ gas, therefore, $\mu_\mathrm{N}$ equals half of the energy of a $\mathrm{N}_2$ molecule ($\mu_\mathrm{N} = \frac{1}{2} \mu_{\mathrm{N}_2}=\mu_\mathrm{N}^{ref}$) and $\mu_\mathrm{B}$ can be calculated as $\mu_\mathrm{B} = \mu_{w\mathrm{BN}}- \mu_\mathrm{N}$, where $\mu_{w\mathrm{BN}}$ is the total energy of pure \wbn\, supercell per formula unit. In nitrogen poor conditions (or boron rich conditions) $\mu_\mathrm{B}=\mu_\mathrm{B}^{ref}$ and $\mu_\mathrm{N}= \mu_{w\mathrm{BN}}- \mu_\mathrm{B}$. It is worth noting that N-rich and B-rich conditions, while not representative of realistic growth environments, serve as theoretical limiting cases.

\begin{figure*}
      \centering
      \includegraphics[scale=0.40]{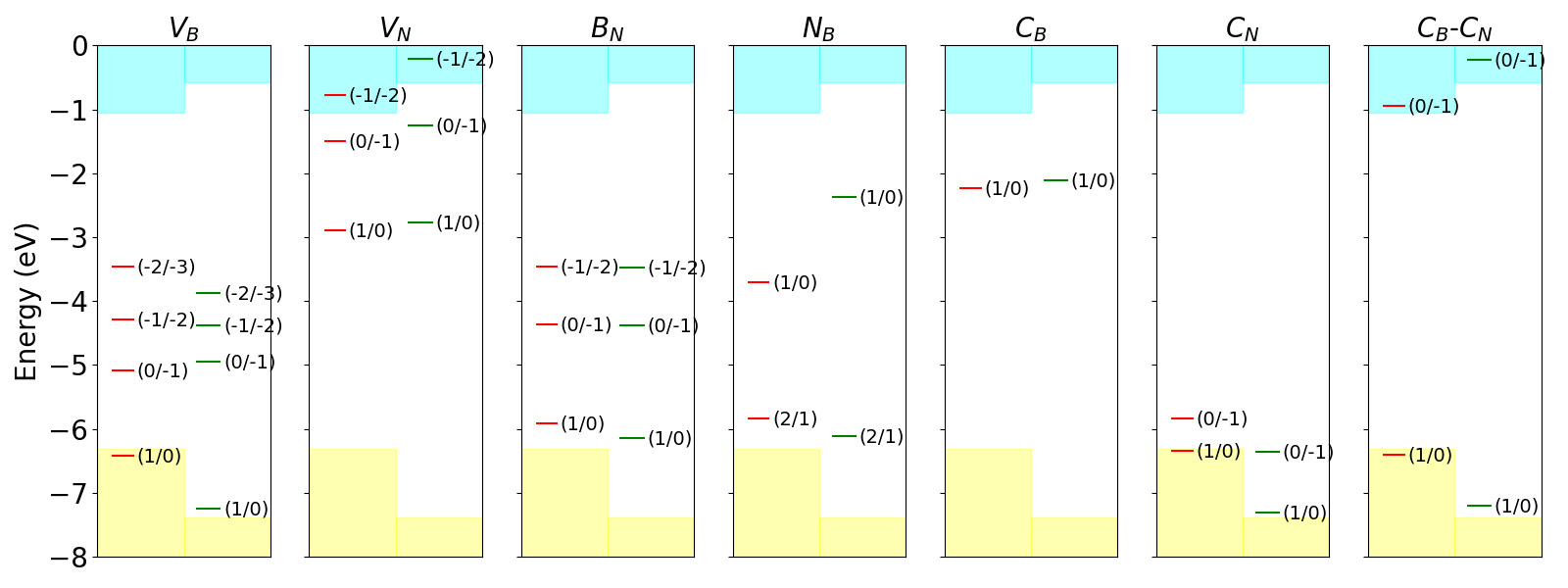}
      \caption{CTLs at PBE (on the left of each plot) and GAU-PBE level (on the right of each plot). The blue area is the conduction band and the yellow area is the valence band of the bulk.\label{fig:CTLs}}.
\end{figure*} 

The charge transition levels (CTL) encompass information about the Fermi energy at which the formation energies of charge states $q$ and $q'$ become equal. They can be obtained as follows with the following expression
\begin{equation}
	\epsilon(q/q')=\frac{E_f[X^{q};E_F=0]-E_f[X^{q'};E_F=0]}{q'-q}.
\end{equation}

In Fig.~\ref{fig:formation_energies_GAU_NB-rich_limits} we show the formation energies at GAU-PBE level in these two limiting cases. The formation energies depend on the nitrogen and boron chemical potentials, except for the carbon dimer. From Eq.\ref{eq:formation_energy} we observe in fact that when two carbon atoms replace two adjacent atoms in the $w$BN lattice the formation energy of the dimer does not depend on the chemical potentials of boron and nitrogen, then it is the same in the two limiting cases. 

The vacancies have a higher formation energy in the limiting case corresponding to the vacant species. In absence of impurities the \vacb\, is the most likely point defect, the other having higher formation energies. The carbon substitutionals have almost the same formation energies (\CsubN\, only slightly higher than the \CsubB) in the N-rich condition due to the total energy difference $E^{def}_{tot}[$\CsubN$]-E^{def}_{tot}[$\CsubB$]$ which almost compensates the difference between $\mu_\mathrm{B} - \mu_\mathrm{N}^{ref}$. On the other hand in the B-rich condition, while the energy difference $E^{def}_{tot}[$\CsubN$]-E^{def}_{tot}[$\CsubB$]$ is still the same, the difference $\mu_\mathrm{B}^{ref} - \mu_\mathrm{N}$ is larger than $\mu_\mathrm{B} - \mu_\mathrm{N}^{ref}$, as a consequence \CsubN\, is favored with respect to \CsubB. The same behavior is observed for the couple of defects \BsubN\, and \NsubB\, in the two limits. 

In Fig.~\ref{fig:CTLs} we report the CTL obtained with PBE and GAU-PBE method and they are discussed in detail for each defect individually in Sec.~\ref{sec:results}.
%{\color{red} QUESTA PARTE È STATA MESSA DOPO We observe that when the Fermi energy is near the valence band maximum, \vacb\, is stable in the positive $q=1$ charge state only at GAU-PBE level. In this case it behaves as a swallow donor. Increasing the Fermi level, it turns into the neutral charge state and at higher Fermi energy values it is stable even at $q=-1$, $q=-2$ and $q=-3$ charge states. \vacb\, is then a very deep acceptor as the transition levels are all found at large energies from the valence band maximum.  
%On the other hand \vacn\, can either be a donor or an acceptor as the two CTLs are found deep in the gap. }

%================================
\section{ Structure and Electronic Properties}
\label{sec:results}
\begin{figure}
	\centering
	\includegraphics[scale=0.36]{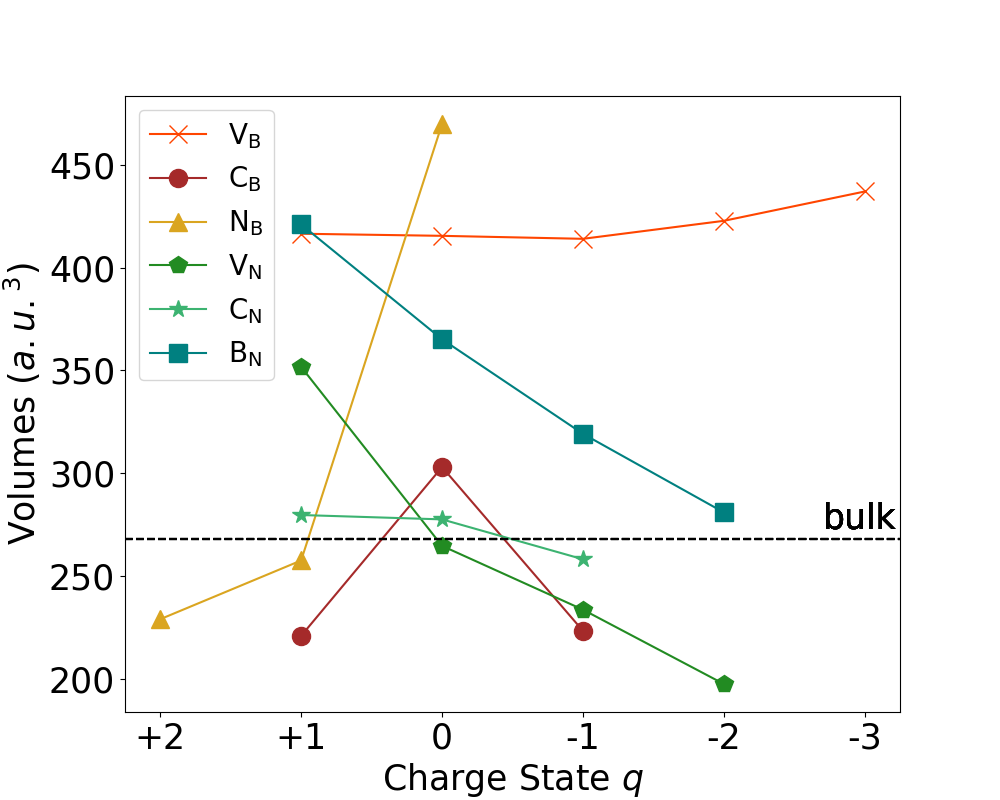}
	\caption{Volumes of the first neighbors tetrahedral cage as function of the charge state for \(\mathrm{V}_{\mathrm{B}}\) (crosses), \(\mathrm{C}_{\mathrm{B}}\) (circles),  \(\mathrm{N}_{\mathrm{B}}\) (triangles), \(\mathrm{V}_{\mathrm{N}}\) (pentagons), \(\mathrm{C}_{\mathrm{N}}\) (stars) and \(\mathrm{B}_{\mathrm{N}}\) (squares). For comparison the black dashed line marks the volume of the tetrahedron formed by the first neighbors of any atom in the pristine crystal.\label{fig:volumes_tetrahedral_cage}}
\end{figure}

%===============================
For each one of the defects and charge states studied, we considered the tetrahedral cage surrounding the defect site, where the vertices are the first four neighboring atoms (as shown in Fig.\ref{fig:AllDef_Struct}(a). The volume of the cage for each defect as a function of the charge state is shown in Fig.\,\ref{fig:volumes_tetrahedral_cage}.
Below we present a defect by defect analysis. For the first, we will give some details. For the others, we will omit them if we consider that they would be repetitive.\\ 

%=====================================================
\emph{Boron Vacancy \(\mathrm{V}_{\mathrm{B}}\)}\\
%=====================================================
If a B atom is removed from the pure supercell, the defective supercell will have three valence electrons less than the pure supercell at \(q\) = 0. Three dangling bonds are created, each filled with a spin up electron. 
%{\color{red}We then predict a magnetisation $\mu_\mathrm{B}$ = 3 magneton Bohr/cell for the $q=0$ charge state, and $\mu_\mathrm{B}=$ 2, 2, 1 and 0 for the \(q=\) 1, -1, -2 and -3 charge states respectively, obtained by removing one electron from the system, and adding one, two and three electrons.}{\color{green}
We then apply Hund's rule and predict a magnetization of $m$ = 3 Bohr magnetons/cell ($m_\mathrm{B}$ per cell) for the $q=0$ charge state, and $m$ = 2, 2, 1 and 0 $m_\mathrm{B}$ per cell for the \(q=\) 1, -1, -2 and -3 charge states respectively, corresponding to removing one electron from the system, and adding one, two and three electrons. 
For each charge state we first relaxed the atomic positions and calculated the magnetization at the PBE level to identify the most likely one. 
%{\color{red}We obtained $m_\mathrm{B}$ = 2, 3, 2, 1 and 0 Bohr magneton /cell for \(q=\) 1, 0, -1, -2 and -3 charge states respectively, as previously predicted. }
We obtained $m$ = 2, 3, 2, 1 and 0 $m_\mathrm{B}$ per cell for \(q=\) 1, 0, -1, -2 and -3 charge states respectively, as previously predicted.

From Fig.\,\ref{fig:volumes_tetrahedral_cage} we can see that the volume of the tetrahedral cage increases as a function of the charge state, and is always larger than the tetrahedron formed by the first neighbors of any atom in the pristine crystal.  
At \(q\) = -3, the first nitrogen atoms surrounding the boron vacancy are much further apart than the original atomic positions, increasing their interdistances by about 10\% if compared to the original ones in the pristine supercell. Before the vacancy is introduced the three electrons of the boron atom are shared with the first three nitrogen atoms on the basal plane. This leads to C$_{3v}$ symmetry and a larger distance between the boron atom and the first nitrogen atom along the z-axis.  In the \(q\) = -3 case, the three additional electrons are captured by the nitrogen atoms on the basal plane, which move away from the vacancy by the same amount as the nitrogen atom along the z-axis (thus conserving the C$_{3v}$ symmetry). When the charge state is progressively reduced the first nitrogen atoms distances gradually decrease starting from that along the z-axis down to those on the basal plane. For this discussion see Fig. S1 of Supplemental Materials.

The relaxed atomic positions were then used to calculate the electronic structure at the GAU-PBE level. 

\begin{figure*}
	\centering
	\includegraphics[scale=0.45]{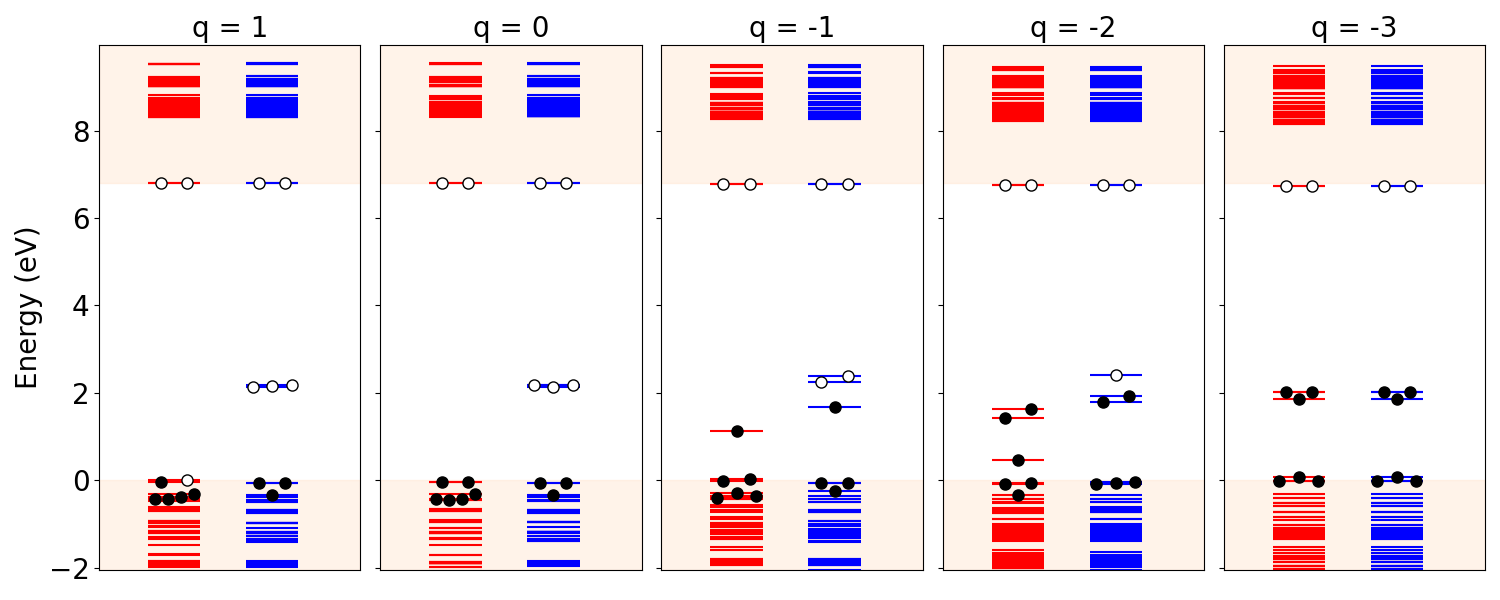}
	\includegraphics[scale=0.45]{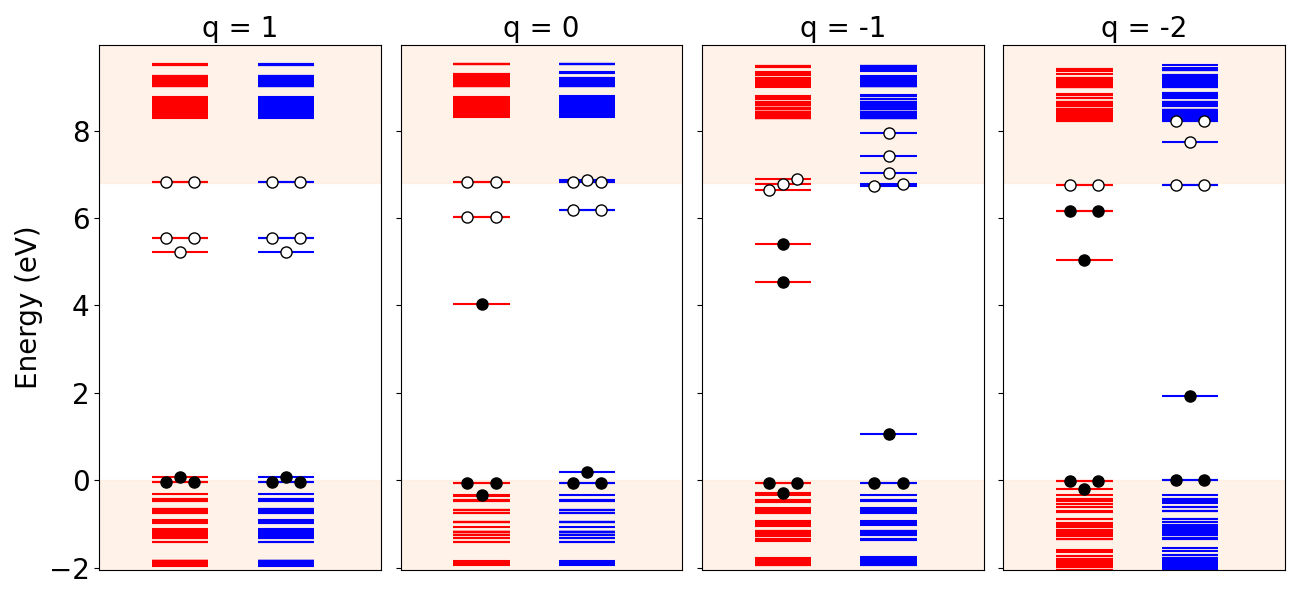}
	\caption{\vacb\, (top panels) and \vacn\, (bottom panels) one-electron levels at GAU-PBE level and different charge states. A black dot represents an occupied state while a blank dot represents an empty state.}
	\label{fig:V_B_EnergyLevels}
\end{figure*}

In the top panels of Fig.\ref{fig:V_B_EnergyLevels} we report the one electron structure for \vacb. It consists of five panels, one for each charge state. The top of the valence band has been set to 0 eV. The pure supercell valence bands below 0 eV and the conduction bands above the gap value of 6.8 eV, are represented by a full colored background. At $q=0$ the lowest energy levels (within 2 eV from the top valence band) in the spin down channel are empty. Since they are two double degenerate levels and one single level very close to each other, we deduce that in the spin up channel there are three uncoupled spin up electrons. The total spin of this defective supercell is then equal to $\nicefrac{3}{2}$ and the total magnetization is 3 $m_{\mathrm{B}}$ per cell. 
At $q=-3$ charge state the magnetization is vanishing, the symmetry in the number of electrons is restored, as adding three electrons to the supercell implies that the spin down levels are again totally filled. 

Moving from $q=-3$ to $q=-2$ or from $q=0$ to $q=1$ charge states, one electron is removed leaving a doubly degenerate state \(e\) with only one electron. In both cases the convergence of the SCF cycles during the relaxation calculations can only be achieved by introducing an effective temperature in the form of a Gaussian smearing (1\(\times10^{-3}\) Ry). We have observed that the occupation of the degenerate \(e\) level does not converge to either 1 or 0, hence one electron is spread equally over two degenerate levels.

The resulting fractional occupation, which indicates a non-physical solution, suggests a Jahn-Teller instability which, not detected by the code, would lift the degeneracy at the cost of distorting the geometry. To find the correct distorted geometry and restore the integer occupancies (either 0 or 1), we moved the atoms around the point defect slightly away from their original atomic positions (less than a tenth of \AA) and allowed the atomic positions to relax. We found a new geometry with a lower total energy configuration and a leverage of the degeneracy in both cases. 

By looking at the CTL in Fig.~\ref{fig:CTLs} we observe that when the Fermi energy is near the valence band maximum, \vacb\, is stable in the positive $q=1$ charge state only at GAU-PBE level. In this case it behaves as a shallow donor. Increasing the Fermi level, it turns into the neutral charge state and at higher Fermi energy values it is stable even at $q=-1$, $q=-2$ and $q=-3$ charge states. \vacb\, is then a very deep acceptor as the transition levels \((0/-1)\), \((-1/-2)\) and \((-2/-3)\) are all found at large energies from the valence band maximum. \\

%======================================
\emph{Nitrogen vacancy \(\mathrm{V}_{\mathrm{N}}\)}\\
%======================================
The removal of a nitrogen atom implies that at charge state \(q=0\) five valence electrons are missing with respect to the original pristine supercell. 
As already done for the \vacb, we consider the resonance between the dangling bonds of the four first neighbors boron atoms surrounding the vacancy site. 
By adding or removing the electrons to and from the boron dangling bond the Hund's rule reads the magnetizations \(m=0, 1, 2, 3\;m_{\mathrm{B}}\) per cell for \(q=1, 0, -1, -2\) respectively. 

As can be seen from Fig.\,\ref{fig:volumes_tetrahedral_cage}, the overall trend shows that the volume shrinks as the number of electrons increases. Similar behavior has previously been observed in oxygen vacancies in ZnO \cite{janotti_van_de_walle_zno} and in nitrogen vacancies in GaN and AlN \cite{Laaksonen_2009}. At \(q\) = 0, elongation of the first boron atom along the z-axis is observed, compensated for by shortening of the remaining boron atoms on the basal plane. Ultimately \(\mathrm{V}_{\mathrm{N}}^{0}\), possesses a volume close to the relaxed pristine material. 
When a nitrogen vacancy is created, the electronic cloud of the boron atoms is attracted to the second-neighbor nitrogen atoms, which are more electronegative. This leads to an expansion of the first boron cage. As soon as electrons are added to the defective supercell, the hybridization of the first neighbors around the point defect increases and the tetrahedral cage contracts inward again. For this discussion see Fig. S2 of Supplemental Materials.

At charge state \(q\) = 1 the electrons are paired, the last occupied levels in the two spin channels are two double degenerate levels $e$ and one single degenerate level $a$ close to the VBM. One single degenerate level and tow double degenerate levels are present well inside the gap below 6 eV, see Fig.\,\ref{fig:V_B_EnergyLevels}. 
At \(q\) = 0, the additional electron occupies a well localized, non-degenerate state inside the gap at around 4 eV above the valence band maximum (VBM). 
The degeneracy of the $e$ states at 6 eV and at the conduction band minimum (CBM) in both spin channels is not disrupted.
This situation will change at charge state \(q\) = -1. The $e$ symmetry of the states described above is lowered to singly degenerate state and the additional
electron will occupy one of them. In the spin down channel these non-degenerate states are shifted to higher energies above the CBM. Finally at \(q\) = -2 a symmetry is partially restored and the non-degenerate states rearrange themselves to form new $e$ levels in spin up channel. 

Calculation of the formation energy found that the two CTL \((+/0)\) and \((0/-)\) are in the gap, so \vacn\, can either be a donor or an acceptor. One more CTL \((-1/-2)\) is found to be in the conduction band. We also simulated the structure for the system \(\mathrm{V}_{\mathrm{N}}^{3-}\) and its formation energy (not reported in Fig.\,\ref{fig:formation_energies_GAU_NB-rich_limits}) which apparently gave rise to a CTL \((+/-3)\) meaning that this defect is a U-center\cite{freysoldt_phys_rev_lett}. 
We consider this configuration physically unrealistic due to significant structural rearrangement, as the crystal structure for \(\mathrm{V}_{\mathrm{N}}^{3-}\) strongly departs from the original system. The forces computed at each SCF step during the relaxation calculation showed a migration of one of the boron first neighbors from its boron site towards the vacancy site. The convergence was not reached until the original nitrogen vacancy evolved into a boron antisite-boron vacancy complex which is a structure totally different from the system under study. Therefore we scrap out the charge state \(q=-3\) from our formation energy and CTL diagrams in Figs.\,\ref{fig:formation_energies_GAU_NB-rich_limits} and \ref{fig:CTLs}. For none of the charge states explored we found a Jahn-Teller distortion.\\ 
%By inspection of the CTLs in Fig.~\ref{fig:CTLs} one can see that \vacn\ can either be a donor or an acceptor as the two CTLs are found deep in the gap.\\

%{\color{red}PUT A FIGURE OF JAHN TELLER DISTORTIONS MS:The problem here is how to relate the symmetry breaking of the crystal with the symmetry breaking of the electronic states. In the case of \(\mathrm{B}_{\mathrm{N}}\), \(\mathrm{N}_{\mathrm{B}}\) and \(\mathrm{V}_{\mathrm{N}}\) it seems that when we add electrons to the highest symmetry configurations, the nuclei configuration symmetry and the symmetry of the electronic states are unrelated. The structure retains the high symmetry (no Jahn-Teller distortion) but the electron freely lower the symmetry and degeneracies are removed when needed. Instead for \(\mathrm{C}_{\mathrm{N}}\), \(\mathrm{C}_{\mathrm{B}}\) and \(\mathrm{V}_{\mathrm{NB}}\) I need to manually break the symmetry of the nuclear configuration to allow the electrons to remove their symmetries. Why?}\\

\begin{figure}
	\centering
	\includegraphics[scale = 0.19]{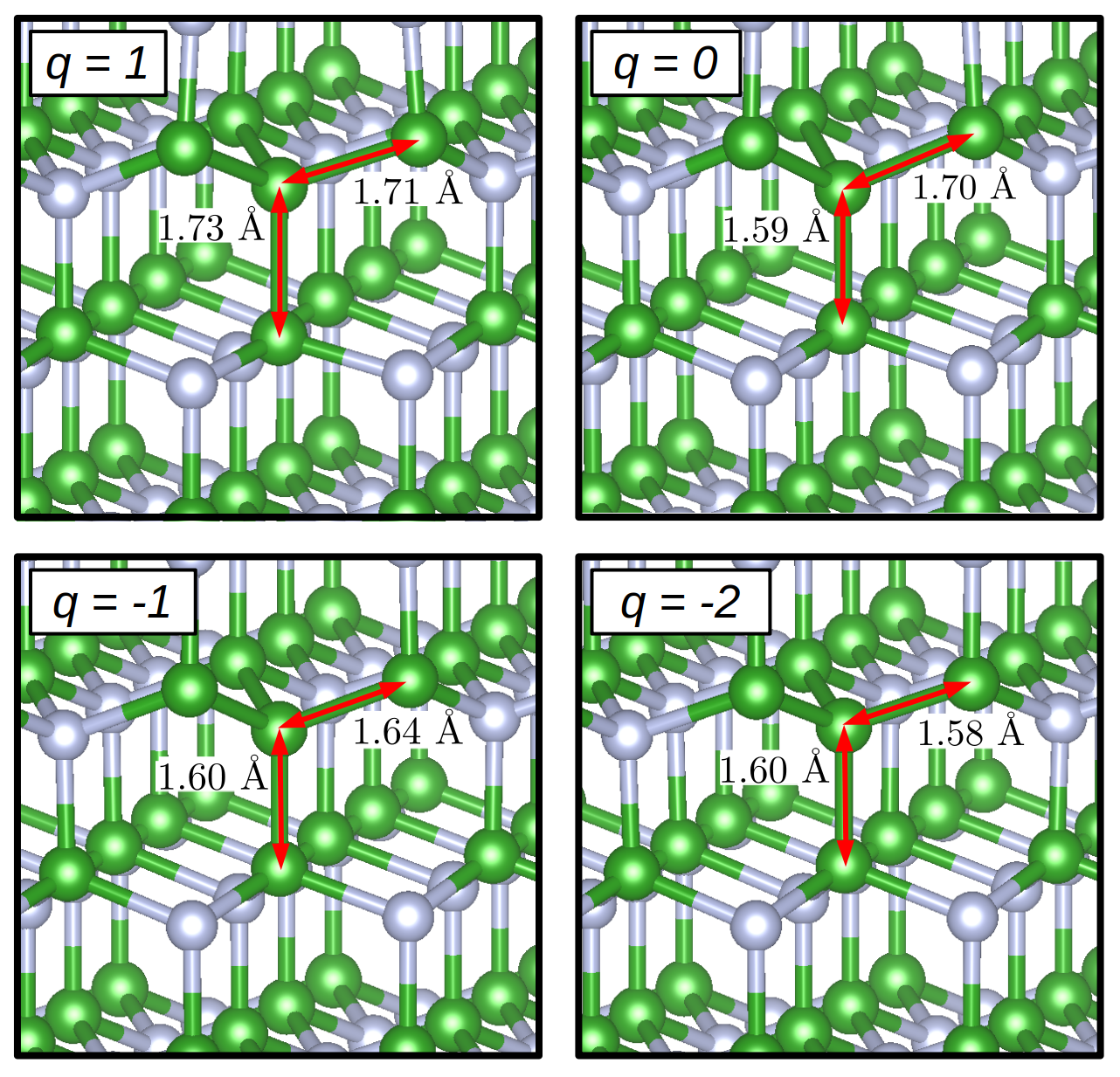}
	\caption{Crystal structure surrounding the boron antisite point defect for four charge states. Distances between the boron antisite and its first neighbors lying on the (0001) plane and along the (0001) axis are highlighted.\label{fig:B_N_struct_dist}}
\end{figure}

%======================================
\emph{Boron antisite \(\mathrm{B}_{\mathrm{N}}\)}\\
%======================================
Fig.\,\ref{fig:B_N_struct_dist} illustrates the crystal structure surrounding the boron antisite for the four charge states discussed below, \(q=1,0,-1\) and \(-2\). 
Replacing a nitrogen atom with a boron atom results in the loss of five valence electrons and the addition of three, making two fewer electrons than the original system when the defect is neutrally charged (\(q=0\)). 

In order to predict the magnetization we considered the \(sp^{3}\) hybridized orbital originated from the boron atom sitting on the defect site and the boron first neighbors, and we filled the orbitals with the electrons belonging to these atoms following the Hund's rule.
%The electron difference between the two spin channels gives the magnetization of the system in Bohr magnetons units \(\mu_{\mathrm{B}}\). 
The predictions, confirmed by DFT calculations, yield \(m = 3, 2, 1 \;m_{\mathrm{B}}\) per cell for \(q = 1, 0, -1\) and vanishing magnetization for \(q=-2\), where the \(sp^{3}\) orbital is completely filled. None of the charge states presents the Jahn-Teller distortion. 

\begin{figure}
	\centering
	\includegraphics[scale = 0.35]{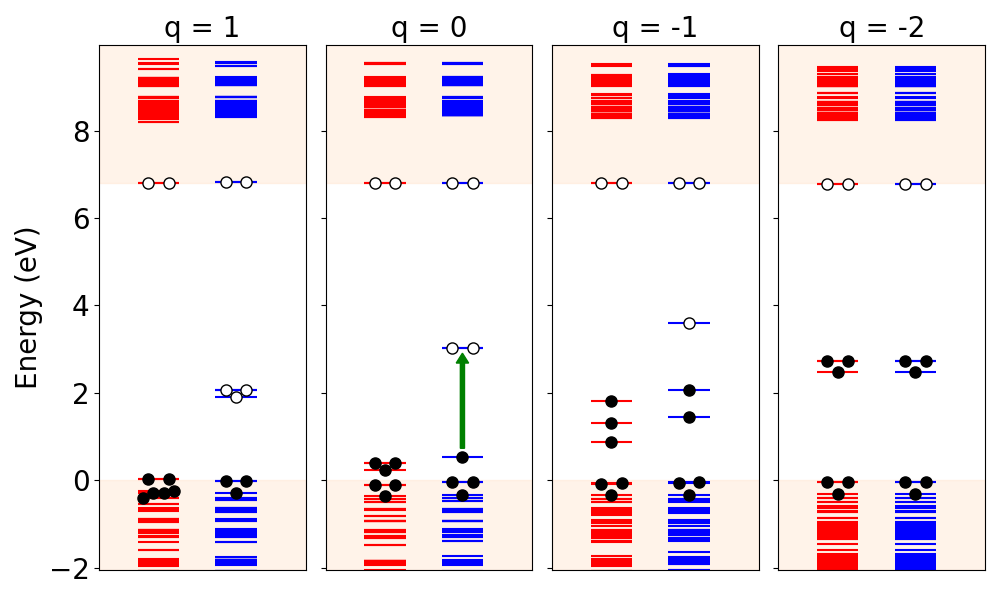}
	\caption{One-electron energy level structure at GAU-PBE level for  \(\mathrm{B}_{\mathrm{N}}\) point defect and for four charge states. A black dot represents an occupied state while a blank dot represents an empty state.\label{fig:B_N_levels}}
\end{figure}
%The one-electron level structures are reported in Fig.\,\ref{fig:B_N_levels} and the distances between the boron antisite and the boron first neighbours are shown in Fig.\,\ref{fig:B_N_struct_dist}.

With reference to Fig.\,\ref{fig:volumes_tetrahedral_cage}, as we increase the number of electrons the volume of the tetrahedral cage decreases always remaining above the bulk value. This trend is similar to the one already described for \(\mathrm{V}_{\mathrm{N}}\). 
%\EC{Even though for \(\mathrm{B}_{\mathrm{N}}\) and \(\mathrm{V}_{\mathrm{N}}\) we studied charge states with different numbers of electrons ??} and we cannot trace a comparison between the two, 
%{\color{red}To draw a comparison between the two at the same charge, in the presence of an electron deficit (i.e. the removal of a nitrogen atom, which has a higher electronegativity than boron), the addition of negative charges increases the hybridisation between the first neighbours of the defect, causing the bond length on the (\(0001\)) plane to decrease.} 
%{\color{green}We draw a comparison between the two. We are in the presence of an electron deficit in both cases since a nitrogen atom has been removed. Since nitrogen has a higher electronegativity than boron, the addition of negative charges increases the hybridization of the first neighbours around \(\mathrm{V}_{\mathrm{N}}\) and \(\mathrm{B}_{\mathrm{N}}\) among themselves rather than the surrounding host environment, thus explaining the common trend of decreasing bond length in the (\(0001\)) plane.} 
%EC: Reminder: Check if in \vacn\, the bond length are similar to \BsubN\, if there is a general trend Rispsota: il trend è molto specifico} 

\begin{figure}
	\centering
	\includegraphics[scale = 0.148]{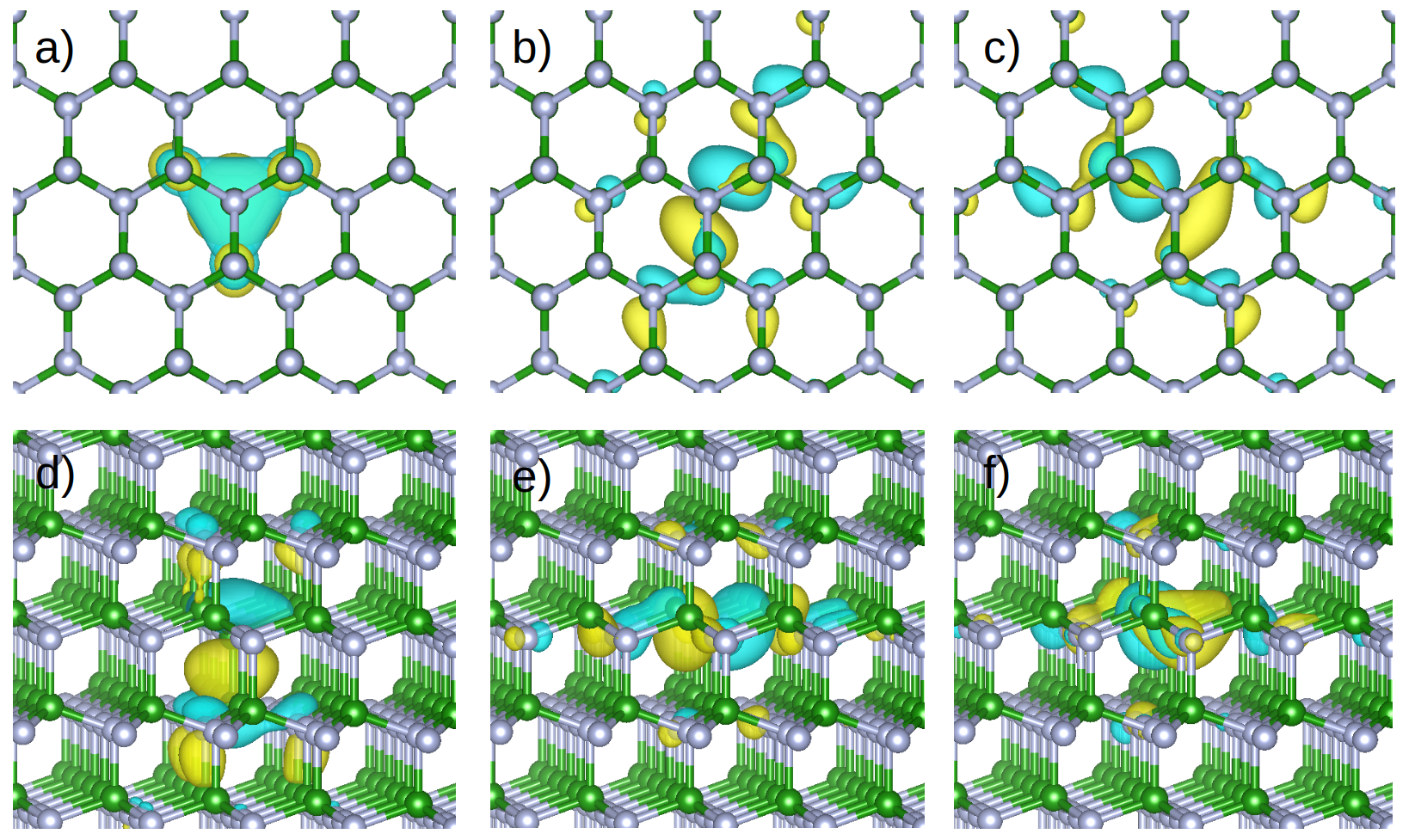}
	\caption{Panels a) and d) \(\left|\psi\right|^{2}\) isosurface for the singly degenerate in-gap state. Panels b, c) and e, f) the doubly degenerate in-gap states of \(\mathrm{B}_{\mathrm{N}}^{0}\), at \(9.09635\times10^{-4}\;a.u.^{-3}\) value for the electronic density. In panels a), b) and c) the reader's eyesight is parallel to the (0001) axis. The shapes do not change sensibly for the same states when the charge is either \(q=1\) or \(q=-2\).\label{fig:B_N_isosurfaces}}
\end{figure}

When \(q=1\), the highest occupied levels are hybridized with the bulk, resulting in delocalised states. As more electrons are added, these levels move towards the center of the gap and become more localized. 
Fig.\,\ref{fig:B_N_isosurfaces} shows the isosurfaces of the singly and doubly degenerate states within the gap. % {\color{red}EC:Spin up or down channel? Which one is plotted? Risposta: tutti e due i canali di spin, quando hanno colori diversi sono solo le fasi positiva e negativa}. 
Considering their degeneracy, shape and the distances between the defect and its nearest neighbors we deduce that \(\mathrm{B}_{\mathrm{N}}^{0}\) has a \(C_{3v}\) symmetry, which is also present in \(\mathrm{B}_{\mathrm{N}}^{+}\) and \(\mathrm{B}_{\mathrm{N}}^{2-}\).
%{\color{red} EC:E'una cosa che diciamo solo a livello della struttura elettronica? Risposta: struttura elettronica e struttura atomica. Non sono sicuro di cosa accada invece per \(\mathrm{B}_{\mathrm{N}}^{-}\)}. 
%JT dinamica viene fuori quando ho gli elettroni spaiati
%Bohr alla meno 3 isosurface
% B_N: Verificare che il blobbo che appare in charge_density q=-1 non appare negli altri stati di carica 
% q=1 Blobbo in tutte le direzioni invece q=-1 no.
% 0.15 as a value 
By contrast, \(\mathrm{B}_{\mathrm{N}}^{-}\) only contains in-gap non-degenerate states, meaning that the \(C_{3v}\) symmetry is lowered. Such lowering in symmetry may be due to a displacement of the charge 
%\EC{da dove a dove? Non lo direi se non ne abbiamo la prova}, 
but it is not a consequence of the Jahn-Teller effect, as no distortion of the atomic positions that would destroy the \(C_{3v}\) symmetry in the crystal was observed. 

Interestingly, we have noticed that the one-electron level structure of \(\mathrm{B}_{\mathrm{N}}^{0}\) is similar to that of the negatively charged nitrogen vacancy complex in diamond\cite{nv_diamon_struct,Gali_review}. Both systems have the same \(C_{3v}\) symmetry and are isoelectronic, and the shape of the isosurfaces of their in-gap states is also similar. 
This suggests that for this particular defect it is possible to optically excite the electron from the singly degenerate in-gap level to the doubly degenerate in-gap level in the minority spin channel (green arrow in Fig.\,\ref{fig:B_N_levels}), thereby constructing the same triplet and singlet states that render the NV complex in diamond useful for quantum sensing purposes. Therefore, the present study suggests that it might be useful to evaluate the Huang-Rhys factor of this defect in order to assess the coherence of the emissions.

The formation energy diagram (see Fig.\,\ref{fig:formation_energies_GAU_NB-rich_limits}) shows that the stable charge states correspond to \(\mathrm{B}_{\mathrm{N}}^{+}\), \(\mathrm{B}_{\mathrm{N}}^{0}\), \(\mathrm{B}_{\mathrm{N}}^{-}\) and \(\mathrm{B}_{\mathrm{N}}^{2-}\) with three CTLs \((+/0)\), \((0/-)\), \((-/2-)\) in the gap. The boron antisite behaves either as an acceptor or donor depending on the position of the Fermi level. \\ 
%at 1.23, 3.01 and 3.90 eV above the VBM, respectively, so that the boron antisite is an amphoteric defect.\\ 
%Amphoteric defects can trap both electrons and holes, acting either as acceptors or donors depending on the position of the Fermi level. 

\begin{figure}
	\centering
	\includegraphics[scale = 0.17]{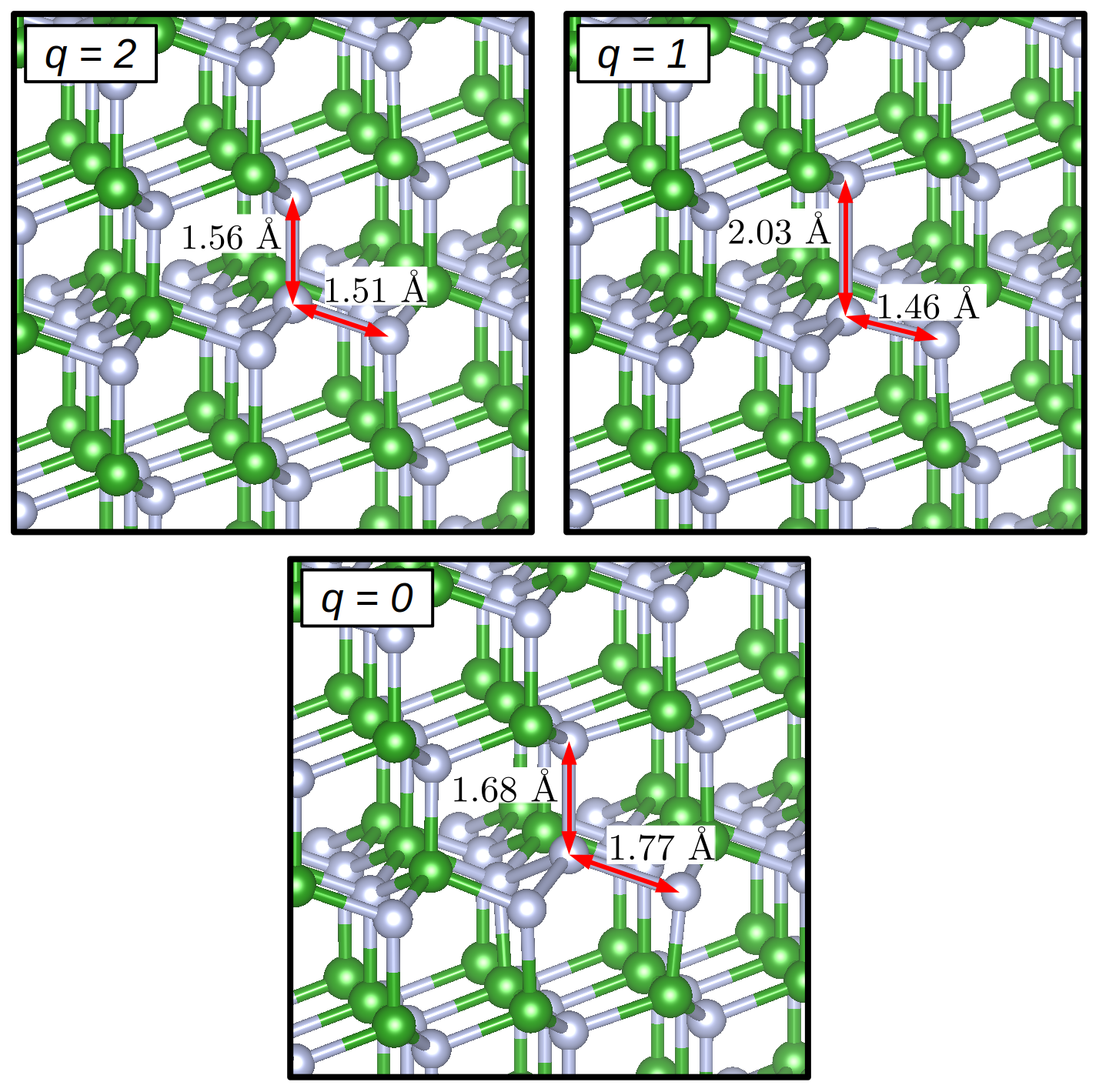}
	\caption{Crystal structure surrounding the nitrogen antisite point defect for three charge states. Distances between the nitrogen antisite and its first neighbors lying on the (0001) plane and along the (0001) axis are highlighted. \label{fig:N_B_struct}}
\end{figure}

%======================================
\emph{Nitrogen antisite \(\mathrm{N}_{\mathrm{B}}\)}\\
%======================================
The crystal structure surrounding the nitrogen antisite is shown in Fig.\,\ref{fig:N_B_struct} for the three charge states discussed below \(q=2,1,0\). 
In this case a nitrogen atom sits on a boron site and it is surrounded by four nitrogen first neighbors. At charge state \(q=0\) three electrons have been removed and five nitrogen electrons added, resulting in two electrons more than in the pristine system.  

The formation energies (see Fig.\,\ref{fig:formation_energies_GAU_NB-rich_limits}) are rather high in both B-rich and N-rich limits, with only \(\mathrm{N}_{\mathrm{B}}^{2+}\) that reaches energies comparable with the one of other defects studied. 
%\MS{for n-type $w$BN}. 
The charge states that are stable in the band gap range are \(q=2\), \(1\) and \(0\) 
%It worths menioning that the relaxation has also been carried out for \(\mathrm{N}_{\mathrm{B}}^{-}\) but SCF convergence could not be achieved whit the hybrid functional. The CTL \((0/-)\) has been calculated only at PBE level and we obtained an energy several eV above the CBM, suggesting that charge state \(q=-1\) is unphysical, therefore it has not been included in the results presented here.   
%and therefore this defect has a donor character, 
with the CTLs \((2+/+)\) and \((+/0)\) far from the bulk conduction band. Thus this defect acts as recombination deep center. 
%\MS{Vogliamo mettere i valori dei CTL in eV?}

\begin{figure}
	\centering
	\includegraphics[scale =0.34]{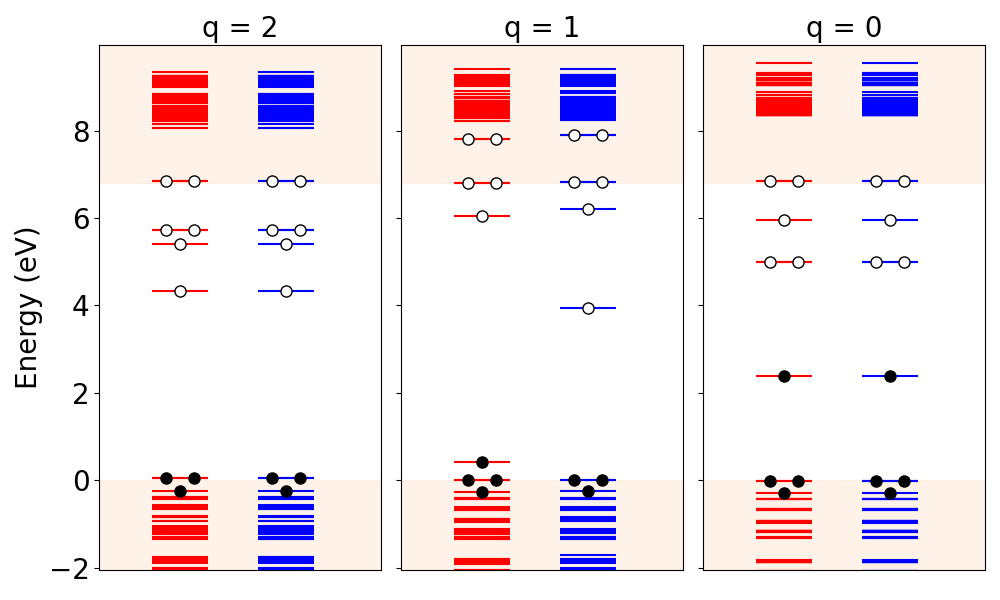}
    \caption{One-electron energy level structure at GAU-PBE level for the \NsubB\, point defect and for three charge states. A black dot represents an occupied state while a blank dot represents an empty state.\label{fig:N_B states}}
\end{figure}

The calculations of the electronic structure provided the magnetizations \(m = 0,1,0 \;m_{\mathrm{B}}\) per cell for \(q=2,1,0\) respectively. These magnetizations cannot be completely ascribed in the model already used in the case of the boron antisite, where the spin configuration has been predicted by applying the Hund's rule on an \(sp^{3}\) orbital. For charge state \(q=0\) the ten electrons involved in each of the N-N bonds do not fit in the four \(sp^{3}\) suborbitals leaving two electrons out, which can rearrange themselves in an antiparallel or parallel configuration. A total energy calculation imposing either a magnetization \(0\) or \(2\;m_{\mathrm{B}}\) per cell respectively, allows to claim that the antiparallel configuration has the lowest total energy. A comparison of the antiparallel spin configuration of \(\mathrm{N}_{\mathrm{B}}^{0}\) with its isoelectronic counterpart \(\mathrm{C}_{\mathrm{B}}^{-}\) will be given when the C substitutionals are discussed. 

By inspection of the relaxed structure (Fig.\,\ref{fig:N_B_struct}) and the one-electron levels structure (Fig.\,\ref{fig:N_B states}) we can argue that the symmetry of the system is \(C_{3v}\) with two singly degenerate (\(a\)) states and a doubly degenerate (\(e\)) state. This symmetry is preserved in all the charge states studied and no Jahn-Teller distortion has been observed.

As shown in Fig.\,\ref{fig:volumes_tetrahedral_cage}, the volume evolution of the first neighbors tetrahedral cage indicates that the crystal structure contracts around the defect with respect to the pristine material for positive charge states \(q = 2, 1\). In contrast, a significant expansion is observed for the neutral charge state. In \(\mathrm{N}_{\mathrm{B}}^{2+}\) two electrons are removed and the system is isoelectronic to the pristine material. Nevertheless, the \(\mathrm{N}_{\mathrm{B}}^{2+}-\mathrm{N}\) bond length along the \(z\)-axis is slightly reduced from 1.58 \AA\, in the bulk material to 1.56 \AA. A larger contraction is also observed for the \(\mathrm{N}_{\mathrm{B}}^{2+}-\mathrm{N}\) bond lengths on the \((0001)\) plane, decreasing from 1.56 \AA\, to 1.51 \AA. This is due to the N atom's different electronegativity compared to its predecessor. 
%replacing the B one. 

\begin{figure}
	\centering
	\includegraphics[scale = 0.2]{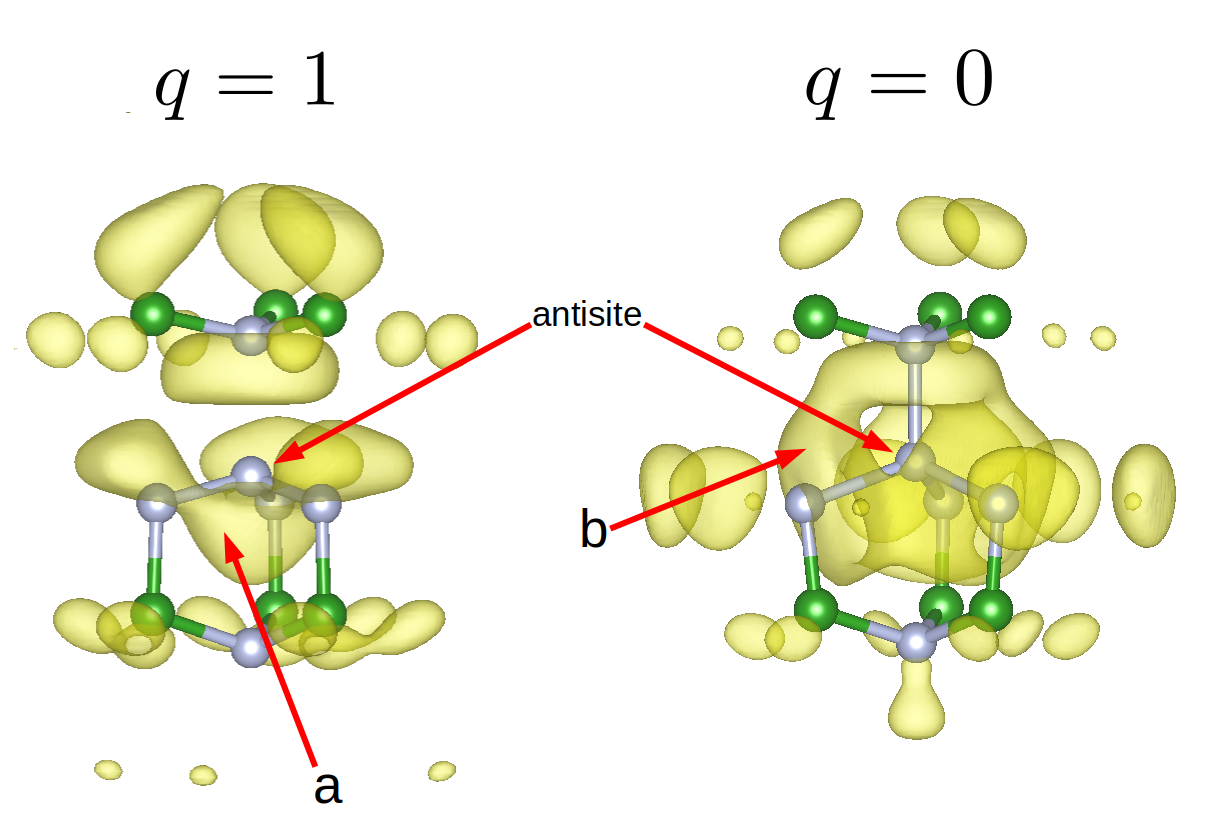}
	\caption{\(\left|\psi\right|^{2}\) isosurface for the highest occupied in-gap state at \(q=1\) and \(q=0\) at \(1.2\times10^{-4}\; a.u.^{-3}\) value for the electronic density. The isosurface region marked as "a" connects only the three defect first neighbors lying on the (0001) plane and the nitrogen antisite. The isosurface marked as "b" involves all four first neighbors but not the defect. For the sake of clarity we show only few atoms around the defect site.\label{fig:N_B isosurface}}
\end{figure}

In \(\mathrm{N}_{\mathrm{B}}^{+2}\) (see the left panel in Fig.\,\ref{fig:N_B states}) the last occupied levels are slightly above the valence band maximum in both spin channels. When an electron is added (\(\mathrm{N}_{\mathrm{B}}^{+}\)) the highest occupied level sits at 0.42 eV above the VBM in the majority spin channel, while a hole occupies the non-degenerate in-gap state in the minority spin channel at 3.94 eV (middle panel in Fig.\,\ref{fig:N_B states}). An inward contraction is also observed in this case. We provide a qualitative explanation for such behavior by examining the shapes of the \(\left|\psi\right|^{2}\) isosurfaces of the occupied in-gap state for \(\mathrm{N}_{\mathrm{B}}^{+}\) (left panel in Fig.\,\ref{fig:N_B isosurface}).
%and \(\mathrm{N}_{\mathrm{B}}^{0}\) (shown in Fig.\,\ref{fig:N_B isosurface}) and considering its bonding-antibonding character. We start from \(q=1\). 
%In \(\mathrm{N}_{\mathrm{B}}^{+}\) the highest occupied level at 0.42 eV above the VBM is in the majority spin channel while a hole sits on the minority spin channel non-degenerate in-gap state at 3.94 eV. 
Here we observe a pronounced axial character of the "a" isosurface. A partial hybridization involves the defect site and the positions of the nitrogen first neighbors on the (0001) plane, and therefore induces an inward contraction around the defect, with the first neighbors-defect distances decreasing from 1.56 \AA\;in the bulk down to 1.46 \AA. This hybridization does not involve, on the contrary, the first neighbor belonging to the (0001) axis above the defect site (upper vertex of the tetrahedral cage) and their distance along the \(z\) axis increases from 1.58 \AA\;in the bulk to 2.03 \AA, thus giving the overall axial character to the isosurface for such state. 

When we move from \(\mathrm{N}_{\mathrm{B}}^{+}\) to \(\mathrm{N}_{\mathrm{B}}^{0}\) the lowest hole is filled and the two highest occupied levels move to 2.39 eV in both spin channels (right panel in Fig.\,\ref{fig:N_B states}). At variance with the previous case, the isosurface "b" (right panel of Fig.\,\ref{fig:N_B isosurface}) has a more isotropic character in the four defect-first neighbors directions and it presents a "shell" shape surrounding the antisite and involving all the four first neighbor nitrogen atoms. The axial character is lost and the first neighbor sitting on the (0001) axis gets closer to the (0001) plane with respect to the previous charge state analyzed. The \(\mathrm{N}_{\mathrm{B}}^{0}-\mathrm{N}\) distance along the $z$ axis decreases to \(1.68\) \AA, while the distance between \(\mathrm{N}_{\mathrm{B}}^{0}\) and the basal first neighbors increases to \(1.77\) \AA. \\ 

%On the other hand it prevents a further hybridization between the first neighbours (which are pulled away by the other bulk atoms) and the defect site itself, which remains more isolated inside the shell. 
 
Carbon is ubiquitously present during growth and it is one of the most common impurities together with oxygen and hydrogen\,\cite{Weston2018}. Given the high formation energy for the native defects studied so far, next we consider the simplest carbon-based point defects in \wbn, to assess their possible role in interfering with an eventual \(p\) or \(n\)-doping for this material.\\

%======================================
\emph{Substitutional Carbon \(\mathrm{C}_{\mathrm{B}}\) and \(\mathrm{C}_{\mathrm{N}}\)}\\
%======================================
\begin{figure}
	\centering
	\includegraphics[scale = 0.34]{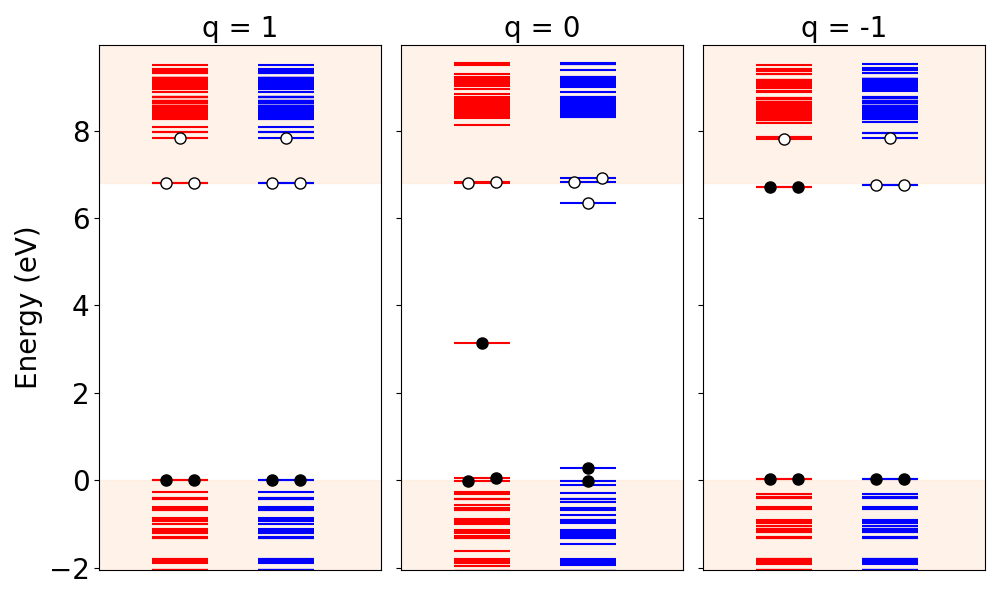}
	\includegraphics[scale = 0.34]{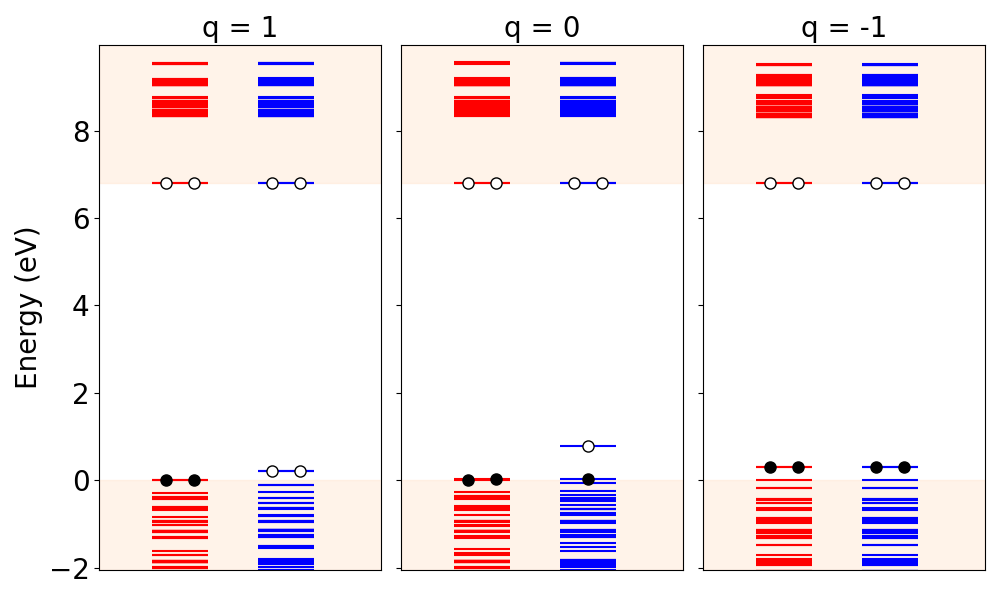}
	\caption{One-electron energy levels at GAU-PBE level for the \(\mathrm{C}_{\mathrm{B}}\) (top panel) and \(\mathrm{C}_{\mathrm{N}}\) (bottom panel) point defects and for three charge states. For charge state \(q=0\) the Jahn-Teller distortion gives origin to the in-gap non degenerate state. A black dot represents an occupied state while a blank
dot represents an empty state.\label{fig:C_B_levels}}
\end{figure}
 In a carbon substitutional defect a carbon atom sits either on a boron site (\(\mathrm{C}_{\mathrm{B}}\)) or a nitrogen site (\(\mathrm{C}_{\mathrm{N}}\)). In both defects we studied the charge states \(q=1, 0, -1\) as they are the most stable inside the band gap. As previously done, the magnetizations for such charge states are obtained by constructing the \(sp^{3}\) hybridized orbitals. 

First we discuss the defect \(\mathrm{C}_{\mathrm{B}}\). Three boron valence electrons are removed and four carbon electrons are added, so that \(\mathrm{C}_{\mathrm{B}}^{0}\) has one electron more with respect to the pristine system. We consider nine electrons (four from the carbon atom and five from the nitrogen first neighbors) to fill the \(sp^{3}\) orbital which can accommodate at most eight electrons, meaning that one electron falls outside the hybridized orbital giving rise for this charge state a magnetization m=1 \(m_{\mathrm{B}}\) per cell. A similar situation occurs for \(\mathrm{C}_{\mathrm{B}}^{-}\) where the electrons that fall outside the \(sp^{3}\) orbital are two. In this case our DFT relaxation predicts for these two excess electrons a parallel spin configuration and the magnetization is 2 \(m_{\mathrm{B}}\) per cell. Finally \(\mathrm{C}_{\mathrm{B}}^{+}\) is isoelectronic to the pristine system, the \(sp^{3}\) orbital is completely filled and the system is non-magnetic. The stable charge states are \(q = 1\) and \(q = 0\), thus this defect has a deep donor character, with a CTL \((+/0)\) at 1.54 below the CBM (5.27 eV above the VBM).

The evolution for the volume for the first neighbors tetrahedral cage shows an irregular behavior, with the charge states \(q=1\) and \(q=-1\) presenting a contraction around the carbon atom and a trend similar to that of \(\mathrm{V}_{\mathrm{B}}\) and \(\mathrm{N}_{\mathrm{B}}\), that is the volume increases with the number of electrons. Nonetheless, this scheme is disrupted by \(\mathrm{C}_{\mathrm{B}}^{0}\) where the Jahn-Teller effect induces a distortion and lowers the symmetry of the crystal geometry (see Fig. S4 of Supplemental Materials). This degeneracy removal can also be inferred by looking at the one-electron level structure of \(\mathrm{C}_{\mathrm{B}}^{0}\) in Fig.\,\ref{fig:C_B_levels} (top panel). 

It is worth noting that \(\mathrm{C}_{\mathrm{B}}^{-}\) is isoelectronic to \(\mathrm{N}_{\mathrm{B}}^{0}\). In both cases there are two electrons that fall outside the hybridized orbital, but the parallel configuration is preferred in \(\mathrm{C}_{\mathrm{B}}^{-}\). The different behavior of the two systems can be qualitatively explained by considering their particular one-electron level structures which are in turn dictated by their symmetry. The level structure of \(\mathrm{N}_{\mathrm{B}}^{0}\) can be obtained by adding two electrons to the lowest unoccupied one-electron levels of \(\mathrm{N}_{\mathrm{B}}^{2+}\) which are singly-degenerate, one in the spin up and the other in the spin down channel. At variance, for \(\mathrm{C}_{\mathrm{B}}^{-}\) the two electrons are added to the lowest unoccupied state of \(\mathrm{C}_{\mathrm{B}}^{+}\), which are doubly degenerate, either of the spin up or the spin down channel, but not the two at the same time, giving a magnetization \(2\;m_{\mathrm{B}}\) per cell. Had these electron half-filled the two doubly degenerate states in spin up and spin down channels, there would have been a degenerate configuration and the system would have experienced a Jahn-Teller distortion. 

We will discuss now the defect \(\mathrm{C}_{\mathrm{N}}\). Five electrons are removed and replaced by four carbon electrons. 
Therefore the neutral charge state \(\mathrm{C}_{\mathrm{N}}^{0}\) hase one electron fewer than the pristine system. The \(sp^{3}\) orbital is left with an unpaired electron, giving rise to a magnetisation of \(1\;m_{\mathrm{B}}\) per cell. When the Fermi level is about 1 eV above the VBM the system can acquire one electron from the environment to complete the octet, resulting in a charge transition to the \(q = -1\) charge state, which has negligible magnetisation. 
%In this case the point defect behaves as a shallow donor. 
Lastly, the state \(\mathrm{C}_{\mathrm{N}}^{+}\) will instead have two fewer electrons than the pristine system, giving a magnetisation of 2 \(m_{\mathrm{B}}\) per cell but the CTL (0/+) falls outside the band gap. Therefore, we did not consider this case.

%, making this defect an acceptor.  
The volume evolution differes between \(\mathrm{C}_{\mathrm{N}}\) and \(\mathrm{C}_{\mathrm{B}}\) cases. 
In the positively charged \(\mathrm{C}_{\mathrm{N}}^{+}\), the atoms on the basal plane (0001) move away from the carbon atom whereas for \(q = -1\), they move towards the substitutional, indicating a higher degree of hybridization, as also observed for the contraction in \(\mathrm{C}_{\mathrm{B}}\). This effect may be due to the greater number of electrons in \(\mathrm{C}_{\mathrm{N}}^{-}\) than is \(\mathrm{C}_{\mathrm{N}}^{+}\). 
Conversely, the parameter dz is always smaller than in the bulk system, indicating anisotropy in the hybridization of the orbitals. 
Looking at the level schemes in Fig.\,\ref{fig:C_B_levels}, we can see that there is a specular behavior between the \(\mathrm{C}_{\mathrm{B}}^{+}\) and \(\mathrm{C}_{\mathrm{N}}^{-}\) systems since they are isoelectronic systems, with levels at resonance with the band edges. Only for the \(\mathrm{C}_{\mathrm{B}}^{0}\) and \(\mathrm{C}_{\mathrm{N}}^{0}\) do some levels sensibly enter the gap.\\

%======================================
\emph{Carbon dimer \(\mathrm{C}_{\mathrm{B}}\mathrm{C}_{\mathrm{N}}\)}\\
%======================================
Finally, we return to the original objective of analysing point defects only, and discuss the carbon dimer, which is a complex formed by two carbon atoms replacing nitrogen and boron atoms that are adjacent to each other. In principle, there are two possible configurations of the carbon dimer: one where the boron and nitrogen sites belong to the basal plane, and one where they belong to the z-axis. 
Here, we only discuss the former case. We studied the charge states \(q=+1\), \(q=0\) and  \(q=-1\). 
The optimised geometric structure of the carbon dimer is shown in Fig. S5 of the Supplementary Materials. The Jahn–Teller effect was not observed for any of the studied charge states.
We observe that the two carbon atoms tend to approach the surrounding nitrogen neighbours more than the first neighbours, with the carbon-carbon and carbon-nitrogen covalent bonds 
being much shorter than the carbon-boron bonds.
The electronegativity of boron is lower than that of nitrogen and carbon, resulting in a charge distribution concentrated on the latter.
The electrically neutral defect is isoelectronic to the original pure material supercell. This is because eight valence electrons are removed (three belonging to the removed boron and five belonging to the removed nitrogen), and these are replaced by eight valence electrons belonging to the two added carbon atoms (four electrons from each carbon atom).
Therefore, we predict that \cdymer\, will be non-magnetic, with the positive and negative charge states having a magnetisation of \(1\;m_{\mathrm{B}}\) per cell. These predictions are confirmed by our DFT relaxations. As expected, the formation energy diagram shows that for Fermi energies within the band gap energy range the neutral defect dominates, with the acceptor and donor character of the two dimer basic constituents, \(\mathrm{C}_{\mathrm{N}}\) and \(\mathrm{C}_{\mathrm{B}}\) respectively, compensating each other, meaning that the carbon dimer is an electrically inactive defect. 
\begin{figure}
	\centering
	\includegraphics[scale =0.34]{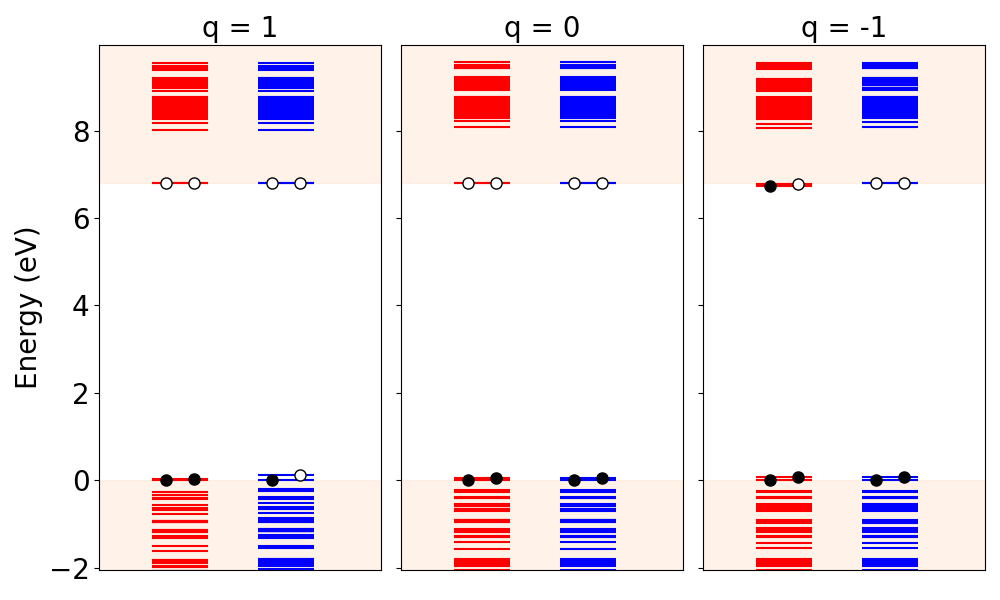}
    \caption{One-electron energy level structure at GAU-PBE level for the carbon dimer and for three charge states. A black dot represents an occupied state while a blank dot represents an empty state.\label{fig:CCstates}}
\end{figure}
%====================
\section{\label{concl}Conclusions}
%====================
In this work, we have conducted a comprehensive first-principles study of native and carbon-based point defects in \wbn, a wide-bandgap semiconductor that has recently been stabilized at ambient pressure. Our analysis includes formation energies with finite-size corrections, charge transition levels, electronic structures, magnetization, and symmetry properties across various charge states, all of which were computed using hybrid functionals.

Among the defects analyzed, the \vacb\, was found to be a deep acceptor with multiple CTLs located well within the band gap. Its electronic structure is characterized by spin-polarized in-gap states and strong Jahn-Teller distortions in certain charge states. The \vacb's rich spin configuration landscape suggests that \vacb\, may host optically active transitions, although its deep level character may limit radiative recombination efficiency. By contrast, the \vacn\, demonstrates both donor- and acceptor-like behavior, with CTLs spanning the band gap. Structural relaxation shows a contraction of the surrounding tetrahedral cage when additional electrons are added, which is consistent with behavior observed in other wide-bandgap semiconductors such as GaN and ZnO. Unlike \vacb, \vacn\, does not exhibit Jahn-Teller distortion, and its charge states evolve smoothly with the position of the Fermi level. 

Regarding the antisites defects, \BsubN\, in its neutral state exhibits structural and electronic properties that closely resemble those of the negatively charged nitrogen vacancy (NV$^-$) center in diamond,
ncluding $C_{3v}$ symmetry and in-gap states. This analogy suggests that \BsubN\, could be a promising candidate for spin-based quantum applications. 
However \NsubB\, has a high formation energy and acts as a deep recombination center. Its charge transition levels indicate both donor and acceptor behavior.

The substitutional carbon defects \CsubB\, and \CsubN\, display distinct electronic characteristics. \CsubB\, predominantly behaves as a deep donor with multiple charge states and significant spin polarization in its negatively charged state. This includes the occurrence of Jahn-Teller distortions in the neutral state. In contrast, \CsubN\, acts primarily as an acceptor with stable charge states confined to the gap and less pronounced magnetic activity. Isoelectronic comparisons between \CsubB$^-$ and \NsubB$^0$ reveal the critical role of local symmetry and orbital occupation in determining the defect magnetization and electronic structure.
FInally we investigated also the \cdymer, which is formed when two adjacent carbon atoms substitute for boron and nitrogen sites, exhibits electrically and magnetically inactive behavior in its neutral state. Similarly, although the carbon dimer defect is electrically neutral and magnetically inactive in its ground state, it may be optically addressable and warrants further investigation for potential quantum emitter behavior.

Overall our results suggest that \wbn\, could be a promising platform for quantum technologies under extreme conditions, offering robust mechanical properties and defect configurations that are favorable for optical and spin manipulation. This study lays the theoretical groundwork for future experimental efforts to engineer and control point defects in \wbn, thereby extending the materials landscape for quantum sensing and communication beyond established hosts such as diamond and silicon carbide. \\

%Our findings suggest that wBN may offer a viable platform for defect-based quantum sensing, expanding the materials landscape beyond established hosts like diamond and silicon carbide, which have already demonstrated exceptional performance in quantum technologies through their NV centers.

%Future experimental validation of the predicted defects and their spin and optical properties could pave the way for realizing wBN-based quantum sensors and emitters.\\

%====================
\section{\label{ackn}Acknowledgments}
%====================
EC and MS acknowledge the support of the French Agence Nationale de la Recherche (ANR) under reference ANR-20-CE47-0009-01-NOTISPERF and ANR-22-CE30-0027-COLIBRI and funding from European Research Council MSCA-ITN TIMES under grant agreement 101118915. Centre de Calcul Intensif d’Aix-Marseille is acknowledged for granting access to its high performance computing resources.  This work was granted access to the HPC/AI resources of TGCC under the allocation 2022-AD010913493 made by GENCI. M.S. and E.C. acknowledge B. Demoulin and  A. Saul for the management of the computer cluster \emph{Rosa}, A. Gali and the members of his group for the inspiring discussions.

\bibliographystyle{apsrev4-1}
\bibliography{defects}% Produces the bibliography via BibTeX.

\begin{thebibliography}{42}%
\makeatletter
\providecommand \@ifxundefined [1]{%
 \@ifx{#1\undefined}
}%
\providecommand \@ifnum [1]{%
 \ifnum #1\expandafter \@firstoftwo
 \else \expandafter \@secondoftwo
 \fi
}%
\providecommand \@ifx [1]{%
 \ifx #1\expandafter \@firstoftwo
 \else \expandafter \@secondoftwo
 \fi
}%
\providecommand \natexlab [1]{#1}%
\providecommand \enquote  [1]{``#1''}%
\providecommand \bibnamefont  [1]{#1}%
\providecommand \bibfnamefont [1]{#1}%
\providecommand \citenamefont [1]{#1}%
\providecommand \href@noop [0]{\@secondoftwo}%
\providecommand \href [0]{\begingroup \@sanitize@url \@href}%
\providecommand \@href[1]{\@@startlink{#1}\@@href}%
\providecommand \@@href[1]{\endgroup#1\@@endlink}%
\providecommand \@sanitize@url [0]{\catcode `\\12\catcode `\$12\catcode
  `\&12\catcode `\#12\catcode `\^12\catcode `\_12\catcode `\%12\relax}%
\providecommand \@@startlink[1]{}%
\providecommand \@@endlink[0]{}%
\providecommand \url  [0]{\begingroup\@sanitize@url \@url }%
\providecommand \@url [1]{\endgroup\@href {#1}{\urlprefix }}%
\providecommand \urlprefix  [0]{URL }%
\providecommand \Eprint [0]{\href }%
\providecommand \doibase [0]{http://dx.doi.org/}%
\providecommand \selectlanguage [0]{\@gobble}%
\providecommand \bibinfo  [0]{\@secondoftwo}%
\providecommand \bibfield  [0]{\@secondoftwo}%
\providecommand \translation [1]{[#1]}%
\providecommand \BibitemOpen [0]{}%
\providecommand \bibitemStop [0]{}%
\providecommand \bibitemNoStop [0]{.\EOS\space}%
\providecommand \EOS [0]{\spacefactor3000\relax}%
\providecommand \BibitemShut  [1]{\csname bibitem#1\endcsname}%
\let\auto@bib@innerbib\@empty
%</preamble>
\bibitem [{\citenamefont {Bundy}\ and\ \citenamefont
  {Wentorf}(2004)}]{bundy_wentorf_wbn_first_synt}%
  \BibitemOpen
  \bibfield  {author} {\bibinfo {author} {\bibfnamefont {F.~P.}\ \bibnamefont
  {Bundy}}\ and\ \bibinfo {author} {\bibfnamefont {J.}~\bibnamefont {Wentorf},
  \bibfnamefont {R.~H.}},\ }\href {\doibase 10.1063/1.1733815} {\bibfield
  {journal} {\bibinfo  {journal} {The Journal of Chemical Physics}\ }\textbf
  {\bibinfo {volume} {38}},\ \bibinfo {pages} {1144} (\bibinfo {year}
  {2004})},\ \Eprint
  {http://arxiv.org/abs/https://pubs.aip.org/aip/jcp/article-pdf/38/5/1144/11273746/1144\_1\_online.pdf}
  {https://pubs.aip.org/aip/jcp/article-pdf/38/5/1144/11273746/1144\_1\_online.pdf}
  \BibitemShut {NoStop}%
\bibitem [{\citenamefont {Segura}\ \emph {et~al.}(2019)\citenamefont {Segura},
  \citenamefont {Cuscó}, \citenamefont {Taniguchi}, \citenamefont {Watanabe},
  \citenamefont {Cassabois}, \citenamefont {Gil},\ and\ \citenamefont
  {Artús}}]{segura}%
  \BibitemOpen
  \bibfield  {author} {\bibinfo {author} {\bibfnamefont {A.}~\bibnamefont
  {Segura}}, \bibinfo {author} {\bibfnamefont {R.}~\bibnamefont {Cuscó}},
  \bibinfo {author} {\bibfnamefont {T.}~\bibnamefont {Taniguchi}}, \bibinfo
  {author} {\bibfnamefont {K.}~\bibnamefont {Watanabe}}, \bibinfo {author}
  {\bibfnamefont {G.}~\bibnamefont {Cassabois}}, \bibinfo {author}
  {\bibfnamefont {B.}~\bibnamefont {Gil}}, \ and\ \bibinfo {author}
  {\bibfnamefont {L.}~\bibnamefont {Artús}},\ }\href {\doibase
  10.1021/acs.jpcc.9b06163} {\bibfield  {journal} {\bibinfo  {journal} {The
  Journal of Physical Chemistry C}\ }\textbf {\bibinfo {volume} {123}},\
  \bibinfo {pages} {20167} (\bibinfo {year} {2019})},\ \Eprint
  {http://arxiv.org/abs/https://doi.org/10.1021/acs.jpcc.9b06163}
  {https://doi.org/10.1021/acs.jpcc.9b06163} \BibitemShut {NoStop}%
\bibitem [{\citenamefont {Solozhenko}\ \emph {et~al.}(1998)\citenamefont
  {Solozhenko}, \citenamefont {Häusermann}, \citenamefont {Mezouar},\ and\
  \citenamefont {Kunz}}]{solozhenko}%
  \BibitemOpen
  \bibfield  {author} {\bibinfo {author} {\bibfnamefont {V.~L.}\ \bibnamefont
  {Solozhenko}}, \bibinfo {author} {\bibfnamefont {D.}~\bibnamefont
  {Häusermann}}, \bibinfo {author} {\bibfnamefont {M.}~\bibnamefont
  {Mezouar}}, \ and\ \bibinfo {author} {\bibfnamefont {M.}~\bibnamefont
  {Kunz}},\ }\href {\doibase 10.1063/1.121186} {\bibfield  {journal} {\bibinfo
  {journal} {Applied Physics Letters}\ }\textbf {\bibinfo {volume} {72}},\
  \bibinfo {pages} {1691} (\bibinfo {year} {1998})},\ \Eprint
  {http://arxiv.org/abs/https://pubs.aip.org/aip/apl/article-pdf/72/14/1691/7806597/1691\_1\_online.pdf}
  {https://pubs.aip.org/aip/apl/article-pdf/72/14/1691/7806597/1691\_1\_online.pdf}
  \BibitemShut {NoStop}%
\bibitem [{\citenamefont {Tani}\ \emph {et~al.}(1975)\citenamefont {Tani},
  \citenamefont {Sōma}, \citenamefont {Sawaoka},\ and\ \citenamefont
  {Saito}}]{Tani_1975}%
  \BibitemOpen
  \bibfield  {author} {\bibinfo {author} {\bibfnamefont {E.}~\bibnamefont
  {Tani}}, \bibinfo {author} {\bibfnamefont {T.}~\bibnamefont {Sōma}},
  \bibinfo {author} {\bibfnamefont {A.}~\bibnamefont {Sawaoka}}, \ and\
  \bibinfo {author} {\bibfnamefont {S.}~\bibnamefont {Saito}},\ }\href
  {\doibase 10.1143/JJAP.14.1605} {\bibfield  {journal} {\bibinfo  {journal}
  {Japanese Journal of Applied Physics}\ }\textbf {\bibinfo {volume} {14}},\
  \bibinfo {pages} {1605} (\bibinfo {year} {1975})}\BibitemShut {NoStop}%
\bibitem [{\citenamefont {Deura}\ \emph {et~al.}(2017)\citenamefont {Deura},
  \citenamefont {Kutsukake}, \citenamefont {Ohno}, \citenamefont {Yonenaga},\
  and\ \citenamefont {Taniguchi}}]{Deura_2017}%
  \BibitemOpen
  \bibfield  {author} {\bibinfo {author} {\bibfnamefont {M.}~\bibnamefont
  {Deura}}, \bibinfo {author} {\bibfnamefont {K.}~\bibnamefont {Kutsukake}},
  \bibinfo {author} {\bibfnamefont {Y.}~\bibnamefont {Ohno}}, \bibinfo {author}
  {\bibfnamefont {I.}~\bibnamefont {Yonenaga}}, \ and\ \bibinfo {author}
  {\bibfnamefont {T.}~\bibnamefont {Taniguchi}},\ }\href {\doibase
  10.7567/JJAP.56.030301} {\bibfield  {journal} {\bibinfo  {journal} {Japanese
  Journal of Applied Physics}\ }\textbf {\bibinfo {volume} {56}},\ \bibinfo
  {pages} {030301} (\bibinfo {year} {2017})}\BibitemShut {NoStop}%
\bibitem [{\citenamefont {Chen}\ \emph {et~al.}(2019)\citenamefont {Chen},
  \citenamefont {Yin}, \citenamefont {Kato}, \citenamefont {Taniguchi},
  \citenamefont {Watanabe}, \citenamefont {Ma}, \citenamefont {Ye},\ and\
  \citenamefont {Ikuhara}}]{Chen2019}%
  \BibitemOpen
  \bibfield  {author} {\bibinfo {author} {\bibfnamefont {C.}~\bibnamefont
  {Chen}}, \bibinfo {author} {\bibfnamefont {D.}~\bibnamefont {Yin}}, \bibinfo
  {author} {\bibfnamefont {T.}~\bibnamefont {Kato}}, \bibinfo {author}
  {\bibfnamefont {T.}~\bibnamefont {Taniguchi}}, \bibinfo {author}
  {\bibfnamefont {K.}~\bibnamefont {Watanabe}}, \bibinfo {author}
  {\bibfnamefont {X.}~\bibnamefont {Ma}}, \bibinfo {author} {\bibfnamefont
  {H.}~\bibnamefont {Ye}}, \ and\ \bibinfo {author} {\bibfnamefont
  {Y.}~\bibnamefont {Ikuhara}},\ }\href {\doibase 10.1073/pnas.1902820116}
  {\bibfield  {journal} {\bibinfo  {journal} {Proceedings of the National
  Academy of Sciences}\ }\textbf {\bibinfo {volume} {116}},\ \bibinfo {pages}
  {11181} (\bibinfo {year} {2019})}\BibitemShut {NoStop}%
\bibitem [{\citenamefont {Yixi}\ \emph {et~al.}(1994)\citenamefont {Yixi},
  \citenamefont {Xin}, \citenamefont {Kun}, \citenamefont {Chaoshu},
  \citenamefont {Zhengfu}, \citenamefont {Junyan}, \citenamefont {Jie},
  \citenamefont {Sheng},\ and\ \citenamefont {Yuanbin}}]{yixi}%
  \BibitemOpen
  \bibfield  {author} {\bibinfo {author} {\bibfnamefont {S.}~\bibnamefont
  {Yixi}}, \bibinfo {author} {\bibfnamefont {J.}~\bibnamefont {Xin}}, \bibinfo
  {author} {\bibfnamefont {W.}~\bibnamefont {Kun}}, \bibinfo {author}
  {\bibfnamefont {S.}~\bibnamefont {Chaoshu}}, \bibinfo {author} {\bibfnamefont
  {H.}~\bibnamefont {Zhengfu}}, \bibinfo {author} {\bibfnamefont
  {S.}~\bibnamefont {Junyan}}, \bibinfo {author} {\bibfnamefont
  {D.}~\bibnamefont {Jie}}, \bibinfo {author} {\bibfnamefont {Z.}~\bibnamefont
  {Sheng}}, \ and\ \bibinfo {author} {\bibfnamefont {C.}~\bibnamefont
  {Yuanbin}},\ }\href {\doibase 10.1103/PhysRevB.50.18637} {\bibfield
  {journal} {\bibinfo  {journal} {Phys. Rev. B}\ }\textbf {\bibinfo {volume}
  {50}},\ \bibinfo {pages} {18637} (\bibinfo {year} {1994})}\BibitemShut
  {NoStop}%
\bibitem [{\citenamefont {Kohn}\ and\ \citenamefont
  {Sham}(1965)}]{ks-equations}%
  \BibitemOpen
  \bibfield  {author} {\bibinfo {author} {\bibfnamefont {W.}~\bibnamefont
  {Kohn}}\ and\ \bibinfo {author} {\bibfnamefont {L.~J.}\ \bibnamefont
  {Sham}},\ }\href {\doibase 10.1103/PhysRev.140.A1133} {\bibfield  {journal}
  {\bibinfo  {journal} {Phys. Rev.}\ }\textbf {\bibinfo {volume} {140}},\
  \bibinfo {pages} {A1133} (\bibinfo {year} {1965})}\BibitemShut {NoStop}%
\bibitem [{\citenamefont {Strinati}(1988)}]{strinati}%
  \BibitemOpen
  \bibfield  {author} {\bibinfo {author} {\bibfnamefont {G.}~\bibnamefont
  {Strinati}},\ }\href@noop {} {\bibfield  {journal} {\bibinfo  {journal} {La
  Rivista del Nuovo Cimento (1978-1999)}\ }\textbf {\bibinfo {volume} {11}},\
  \bibinfo {pages} {1} (\bibinfo {year} {1988})}\BibitemShut {NoStop}%
\bibitem [{\citenamefont {Hedin}(1965)}]{hedin}%
  \BibitemOpen
  \bibfield  {author} {\bibinfo {author} {\bibfnamefont {L.}~\bibnamefont
  {Hedin}},\ }\href {\doibase 10.1103/PhysRev.139.A796} {\bibfield  {journal}
  {\bibinfo  {journal} {Phys. Rev.}\ }\textbf {\bibinfo {volume} {139}},\
  \bibinfo {pages} {A796} (\bibinfo {year} {1965})}\BibitemShut {NoStop}%
\bibitem [{\citenamefont {Onida}\ \emph {et~al.}(2002)\citenamefont {Onida},
  \citenamefont {Reining},\ and\ \citenamefont {Rubio}}]{onida_reining_rubio}%
  \BibitemOpen
  \bibfield  {author} {\bibinfo {author} {\bibfnamefont {G.}~\bibnamefont
  {Onida}}, \bibinfo {author} {\bibfnamefont {L.}~\bibnamefont {Reining}}, \
  and\ \bibinfo {author} {\bibfnamefont {A.}~\bibnamefont {Rubio}},\ }\href
  {\doibase 10.1103/RevModPhys.74.601} {\bibfield  {journal} {\bibinfo
  {journal} {Rev. Mod. Phys.}\ }\textbf {\bibinfo {volume} {74}},\ \bibinfo
  {pages} {601} (\bibinfo {year} {2002})}\BibitemShut {NoStop}%
\bibitem [{\citenamefont {Silvetti}\ \emph {et~al.}(2023)\citenamefont
  {Silvetti}, \citenamefont {Attaccalite},\ and\ \citenamefont
  {Cannuccia}}]{silvetti23}%
  \BibitemOpen
  \bibfield  {author} {\bibinfo {author} {\bibfnamefont {M.}~\bibnamefont
  {Silvetti}}, \bibinfo {author} {\bibfnamefont {C.}~\bibnamefont
  {Attaccalite}}, \ and\ \bibinfo {author} {\bibfnamefont {E.}~\bibnamefont
  {Cannuccia}},\ }\href {\doibase 10.1103/PhysRevMaterials.7.055201} {\bibfield
   {journal} {\bibinfo  {journal} {Phys. Rev. Mater.}\ }\textbf {\bibinfo
  {volume} {7}},\ \bibinfo {pages} {055201} (\bibinfo {year}
  {2023})}\BibitemShut {NoStop}%
\bibitem [{\citenamefont {Christensen}\ and\ \citenamefont
  {Gorczyca}(1994)}]{christensen}%
  \BibitemOpen
  \bibfield  {author} {\bibinfo {author} {\bibfnamefont {N.~E.}\ \bibnamefont
  {Christensen}}\ and\ \bibinfo {author} {\bibfnamefont {I.}~\bibnamefont
  {Gorczyca}},\ }\href {\doibase 10.1103/PhysRevB.50.4397} {\bibfield
  {journal} {\bibinfo  {journal} {Phys. Rev. B}\ }\textbf {\bibinfo {volume}
  {50}},\ \bibinfo {pages} {4397} (\bibinfo {year} {1994})}\BibitemShut
  {NoStop}%
\bibitem [{\citenamefont {Cappellini}\ \emph {et~al.}(2000)\citenamefont
  {Cappellini}, \citenamefont {Satta}, \citenamefont {Tenelsen},\ and\
  \citenamefont {Bechstedt}}]{cappellini}%
  \BibitemOpen
  \bibfield  {author} {\bibinfo {author} {\bibfnamefont {G.}~\bibnamefont
  {Cappellini}}, \bibinfo {author} {\bibfnamefont {G.}~\bibnamefont {Satta}},
  \bibinfo {author} {\bibfnamefont {K.}~\bibnamefont {Tenelsen}}, \ and\
  \bibinfo {author} {\bibfnamefont {F.}~\bibnamefont {Bechstedt}},\ }\href
  {\doibase
  https://doi.org/10.1002/(SICI)1521-3951(200002)217:2<861::AID-PSSB861>3.0.CO;2-H}
  {\bibfield  {journal} {\bibinfo  {journal} {physica status solidi (b)}\
  }\textbf {\bibinfo {volume} {217}},\ \bibinfo {pages} {861} (\bibinfo {year}
  {2000})}\BibitemShut {NoStop}%
\bibitem [{\citenamefont {Weston}\ \emph {et~al.}(2018)\citenamefont {Weston},
  \citenamefont {Wickramaratne}, \citenamefont {Mackoit}, \citenamefont
  {Alkauskas},\ and\ \citenamefont {Van~de Walle}}]{Weston2018}%
  \BibitemOpen
  \bibfield  {author} {\bibinfo {author} {\bibfnamefont {L.}~\bibnamefont
  {Weston}}, \bibinfo {author} {\bibfnamefont {D.}~\bibnamefont
  {Wickramaratne}}, \bibinfo {author} {\bibfnamefont {M.}~\bibnamefont
  {Mackoit}}, \bibinfo {author} {\bibfnamefont {A.}~\bibnamefont {Alkauskas}},
  \ and\ \bibinfo {author} {\bibfnamefont {C.~G.}\ \bibnamefont {Van~de
  Walle}},\ }\href {\doibase 10.1103/PhysRevB.97.214104} {\bibfield  {journal}
  {\bibinfo  {journal} {Phys. Rev. B}\ }\textbf {\bibinfo {volume} {97}},\
  \bibinfo {pages} {214104} (\bibinfo {year} {2018})}\BibitemShut {NoStop}%
\bibitem [{\citenamefont {Orellana}\ and\ \citenamefont
  {Chacham}(2001)}]{orellana2001stability}%
  \BibitemOpen
  \bibfield  {author} {\bibinfo {author} {\bibfnamefont {W.}~\bibnamefont
  {Orellana}}\ and\ \bibinfo {author} {\bibfnamefont {H.}~\bibnamefont
  {Chacham}},\ }\href@noop {} {\bibfield  {journal} {\bibinfo  {journal}
  {Physical Review B}\ }\textbf {\bibinfo {volume} {63}},\ \bibinfo {pages}
  {125205} (\bibinfo {year} {2001})}\BibitemShut {NoStop}%
\bibitem [{\citenamefont {Tararan}\ \emph {et~al.}(2018)\citenamefont
  {Tararan}, \citenamefont {di~Sabatino}, \citenamefont {Gatti}, \citenamefont
  {Taniguchi}, \citenamefont {Watanabe}, \citenamefont {Reining}, \citenamefont
  {Tizei}, \citenamefont {Kociak},\ and\ \citenamefont
  {Zobelli}}]{tararan_zobelli}%
  \BibitemOpen
  \bibfield  {author} {\bibinfo {author} {\bibfnamefont {A.}~\bibnamefont
  {Tararan}}, \bibinfo {author} {\bibfnamefont {S.}~\bibnamefont
  {di~Sabatino}}, \bibinfo {author} {\bibfnamefont {M.}~\bibnamefont {Gatti}},
  \bibinfo {author} {\bibfnamefont {T.}~\bibnamefont {Taniguchi}}, \bibinfo
  {author} {\bibfnamefont {K.}~\bibnamefont {Watanabe}}, \bibinfo {author}
  {\bibfnamefont {L.}~\bibnamefont {Reining}}, \bibinfo {author} {\bibfnamefont
  {L.~H.~G.}\ \bibnamefont {Tizei}}, \bibinfo {author} {\bibfnamefont
  {M.}~\bibnamefont {Kociak}}, \ and\ \bibinfo {author} {\bibfnamefont
  {A.}~\bibnamefont {Zobelli}},\ }\href {\doibase 10.1103/PhysRevB.98.094106}
  {\bibfield  {journal} {\bibinfo  {journal} {Phys. Rev. B}\ }\textbf {\bibinfo
  {volume} {98}},\ \bibinfo {pages} {094106} (\bibinfo {year}
  {2018})}\BibitemShut {NoStop}%
\bibitem [{\citenamefont {Mart\'{\i}nez}\ \emph {et~al.}(2016)\citenamefont
  {Mart\'{\i}nez}, \citenamefont {Pelini}, \citenamefont {Waselowski},
  \citenamefont {Maze}, \citenamefont {Gil}, \citenamefont {Cassabois},\ and\
  \citenamefont {Jacques}}]{Martinez2016}%
  \BibitemOpen
  \bibfield  {author} {\bibinfo {author} {\bibfnamefont {L.~J.}\ \bibnamefont
  {Mart\'{\i}nez}}, \bibinfo {author} {\bibfnamefont {T.}~\bibnamefont
  {Pelini}}, \bibinfo {author} {\bibfnamefont {V.}~\bibnamefont {Waselowski}},
  \bibinfo {author} {\bibfnamefont {J.~R.}\ \bibnamefont {Maze}}, \bibinfo
  {author} {\bibfnamefont {B.}~\bibnamefont {Gil}}, \bibinfo {author}
  {\bibfnamefont {G.}~\bibnamefont {Cassabois}}, \ and\ \bibinfo {author}
  {\bibfnamefont {V.}~\bibnamefont {Jacques}},\ }\href {\doibase
  10.1103/PhysRevB.94.121405} {\bibfield  {journal} {\bibinfo  {journal} {Phys.
  Rev. B}\ }\textbf {\bibinfo {volume} {94}},\ \bibinfo {pages} {121405}
  (\bibinfo {year} {2016})}\BibitemShut {NoStop}%
\bibitem [{\citenamefont {Bourrellier}\ \emph {et~al.}(2016)\citenamefont
  {Bourrellier}, \citenamefont {Meuret}, \citenamefont {Tararan}, \citenamefont
  {St{\'e}phan}, \citenamefont {Kociak}, \citenamefont {Tizei},\ and\
  \citenamefont {Zobelli}}]{Bourrellier2016}%
  \BibitemOpen
  \bibfield  {author} {\bibinfo {author} {\bibfnamefont {R.}~\bibnamefont
  {Bourrellier}}, \bibinfo {author} {\bibfnamefont {S.}~\bibnamefont {Meuret}},
  \bibinfo {author} {\bibfnamefont {A.}~\bibnamefont {Tararan}}, \bibinfo
  {author} {\bibfnamefont {O.}~\bibnamefont {St{\'e}phan}}, \bibinfo {author}
  {\bibfnamefont {M.}~\bibnamefont {Kociak}}, \bibinfo {author} {\bibfnamefont
  {L.~H.~G.}\ \bibnamefont {Tizei}}, \ and\ \bibinfo {author} {\bibfnamefont
  {A.}~\bibnamefont {Zobelli}},\ }\href {\doibase 10.1021/acs.nanolett.6b01368}
  {\bibfield  {journal} {\bibinfo  {journal} {Nano Letters}\ }\textbf {\bibinfo
  {volume} {16}},\ \bibinfo {pages} {4317} (\bibinfo {year} {2016})},\ \bibinfo
  {note} {pMID: 27299915},\ \Eprint
  {http://arxiv.org/abs/https://doi.org/10.1021/acs.nanolett.6b01368}
  {https://doi.org/10.1021/acs.nanolett.6b01368} \BibitemShut {NoStop}%
\bibitem [{\citenamefont {Mackoit-Sinkevičienė}\ \emph
  {et~al.}(2019)\citenamefont {Mackoit-Sinkevičienė}, \citenamefont
  {Maciaszek}, \citenamefont {Van~de Walle},\ and\ \citenamefont
  {Alkauskas}}]{mackoit2019}%
  \BibitemOpen
  \bibfield  {author} {\bibinfo {author} {\bibfnamefont {M.}~\bibnamefont
  {Mackoit-Sinkevičienė}}, \bibinfo {author} {\bibfnamefont {M.}~\bibnamefont
  {Maciaszek}}, \bibinfo {author} {\bibfnamefont {C.~G.}\ \bibnamefont {Van~de
  Walle}}, \ and\ \bibinfo {author} {\bibfnamefont {A.}~\bibnamefont
  {Alkauskas}},\ }\href {\doibase 10.1063/1.5124153} {\bibfield  {journal}
  {\bibinfo  {journal} {Applied Physics Letters}\ }\textbf {\bibinfo {volume}
  {115}},\ \bibinfo {pages} {212101} (\bibinfo {year} {2019})},\ \Eprint
  {http://arxiv.org/abs/https://pubs.aip.org/aip/apl/article-pdf/doi/10.1063/1.5124153/13286483/212101\_1\_online.pdf}
  {https://pubs.aip.org/aip/apl/article-pdf/doi/10.1063/1.5124153/13286483/212101\_1\_online.pdf}
  \BibitemShut {NoStop}%
\bibitem [{\citenamefont {Nguyen}\ \emph {et~al.}(2024)\citenamefont {Nguyen},
  \citenamefont {Dang}, \citenamefont {Pham},\ and\ \citenamefont
  {Nghiem}}]{nguyen2024computationally}%
  \BibitemOpen
  \bibfield  {author} {\bibinfo {author} {\bibfnamefont {N.~L.}\ \bibnamefont
  {Nguyen}}, \bibinfo {author} {\bibfnamefont {H.~T.}\ \bibnamefont {Dang}},
  \bibinfo {author} {\bibfnamefont {T.~L.}\ \bibnamefont {Pham}}, \ and\
  \bibinfo {author} {\bibfnamefont {T.~M.~H.}\ \bibnamefont {Nghiem}},\
  }\href@noop {} {\bibfield  {journal} {\bibinfo  {journal} {arXiv preprint
  arXiv:2402.08464}\ } (\bibinfo {year} {2024})}\BibitemShut {NoStop}%
\bibitem [{\citenamefont {Albe}(1997)}]{albe}%
  \BibitemOpen
  \bibfield  {author} {\bibinfo {author} {\bibfnamefont {K.}~\bibnamefont
  {Albe}},\ }\href {\doibase 10.1103/PhysRevB.55.6203} {\bibfield  {journal}
  {\bibinfo  {journal} {Phys. Rev. B}\ }\textbf {\bibinfo {volume} {55}},\
  \bibinfo {pages} {6203} (\bibinfo {year} {1997})}\BibitemShut {NoStop}%
\bibitem [{\citenamefont {Orellana}\ and\ \citenamefont
  {Chacham}(2000)}]{orellana}%
  \BibitemOpen
  \bibfield  {author} {\bibinfo {author} {\bibfnamefont {W.}~\bibnamefont
  {Orellana}}\ and\ \bibinfo {author} {\bibfnamefont {H.}~\bibnamefont
  {Chacham}},\ }\href {\doibase 10.1103/PhysRevB.62.10135} {\bibfield
  {journal} {\bibinfo  {journal} {Phys. Rev. B}\ }\textbf {\bibinfo {volume}
  {62}},\ \bibinfo {pages} {10135} (\bibinfo {year} {2000})}\BibitemShut
  {NoStop}%
\bibitem [{\citenamefont {Giannozzi}\ \emph {et~al.}(2017)\citenamefont
  {Giannozzi}, \citenamefont {Andreussi}, \citenamefont {Brumme}, \citenamefont
  {Bunau}, \citenamefont {Nardelli}, \citenamefont {Calandra}, \citenamefont
  {Car}, \citenamefont {Cavazzoni}, \citenamefont {Ceresoli}, \citenamefont
  {Cococcioni}, \citenamefont {Colonna}, \citenamefont {Carnimeo},
  \citenamefont {Corso}, \citenamefont {de~Gironcoli}, \citenamefont {Delugas},
  \citenamefont {DiStasio}, \citenamefont {Ferretti}, \citenamefont {Floris},
  \citenamefont {Fratesi}, \citenamefont {Fugallo}, \citenamefont {Gebauer},
  \citenamefont {Gerstmann}, \citenamefont {Giustino}, \citenamefont {Gorni},
  \citenamefont {Jia}, \citenamefont {Kawamura}, \citenamefont {Ko},
  \citenamefont {Kokalj}, \citenamefont {Küçükbenli}, \citenamefont
  {Lazzeri}, \citenamefont {Marsili}, \citenamefont {Marzari}, \citenamefont
  {Mauri}, \citenamefont {Nguyen}, \citenamefont {Nguyen}, \citenamefont {de-la
  Roza}, \citenamefont {Paulatto}, \citenamefont {Poncé}, \citenamefont
  {Rocca}, \citenamefont {Sabatini}, \citenamefont {Santra}, \citenamefont
  {Schlipf}, \citenamefont {Seitsonen}, \citenamefont {Smogunov}, \citenamefont
  {Timrov}, \citenamefont {Thonhauser}, \citenamefont {Umari}, \citenamefont
  {Vast}, \citenamefont {Wu},\ and\ \citenamefont {Baroni}}]{Giannozzi_2017}%
  \BibitemOpen
  \bibfield  {author} {\bibinfo {author} {\bibfnamefont {P.}~\bibnamefont
  {Giannozzi}}, \bibinfo {author} {\bibfnamefont {O.}~\bibnamefont
  {Andreussi}}, \bibinfo {author} {\bibfnamefont {T.}~\bibnamefont {Brumme}},
  \bibinfo {author} {\bibfnamefont {O.}~\bibnamefont {Bunau}}, \bibinfo
  {author} {\bibfnamefont {M.~B.}\ \bibnamefont {Nardelli}}, \bibinfo {author}
  {\bibfnamefont {M.}~\bibnamefont {Calandra}}, \bibinfo {author}
  {\bibfnamefont {R.}~\bibnamefont {Car}}, \bibinfo {author} {\bibfnamefont
  {C.}~\bibnamefont {Cavazzoni}}, \bibinfo {author} {\bibfnamefont
  {D.}~\bibnamefont {Ceresoli}}, \bibinfo {author} {\bibfnamefont
  {M.}~\bibnamefont {Cococcioni}}, \bibinfo {author} {\bibfnamefont
  {N.}~\bibnamefont {Colonna}}, \bibinfo {author} {\bibfnamefont
  {I.}~\bibnamefont {Carnimeo}}, \bibinfo {author} {\bibfnamefont {A.~D.}\
  \bibnamefont {Corso}}, \bibinfo {author} {\bibfnamefont {S.}~\bibnamefont
  {de~Gironcoli}}, \bibinfo {author} {\bibfnamefont {P.}~\bibnamefont
  {Delugas}}, \bibinfo {author} {\bibfnamefont {R.~A.}\ \bibnamefont
  {DiStasio}}, \bibinfo {author} {\bibfnamefont {A.}~\bibnamefont {Ferretti}},
  \bibinfo {author} {\bibfnamefont {A.}~\bibnamefont {Floris}}, \bibinfo
  {author} {\bibfnamefont {G.}~\bibnamefont {Fratesi}}, \bibinfo {author}
  {\bibfnamefont {G.}~\bibnamefont {Fugallo}}, \bibinfo {author} {\bibfnamefont
  {R.}~\bibnamefont {Gebauer}}, \bibinfo {author} {\bibfnamefont
  {U.}~\bibnamefont {Gerstmann}}, \bibinfo {author} {\bibfnamefont
  {F.}~\bibnamefont {Giustino}}, \bibinfo {author} {\bibfnamefont
  {T.}~\bibnamefont {Gorni}}, \bibinfo {author} {\bibfnamefont
  {J.}~\bibnamefont {Jia}}, \bibinfo {author} {\bibfnamefont {M.}~\bibnamefont
  {Kawamura}}, \bibinfo {author} {\bibfnamefont {H.-Y.}\ \bibnamefont {Ko}},
  \bibinfo {author} {\bibfnamefont {A.}~\bibnamefont {Kokalj}}, \bibinfo
  {author} {\bibfnamefont {E.}~\bibnamefont {Küçükbenli}}, \bibinfo {author}
  {\bibfnamefont {M.}~\bibnamefont {Lazzeri}}, \bibinfo {author} {\bibfnamefont
  {M.}~\bibnamefont {Marsili}}, \bibinfo {author} {\bibfnamefont
  {N.}~\bibnamefont {Marzari}}, \bibinfo {author} {\bibfnamefont
  {F.}~\bibnamefont {Mauri}}, \bibinfo {author} {\bibfnamefont {N.~L.}\
  \bibnamefont {Nguyen}}, \bibinfo {author} {\bibfnamefont {H.-V.}\
  \bibnamefont {Nguyen}}, \bibinfo {author} {\bibfnamefont {A.~O.}\
  \bibnamefont {de-la Roza}}, \bibinfo {author} {\bibfnamefont
  {L.}~\bibnamefont {Paulatto}}, \bibinfo {author} {\bibfnamefont
  {S.}~\bibnamefont {Poncé}}, \bibinfo {author} {\bibfnamefont
  {D.}~\bibnamefont {Rocca}}, \bibinfo {author} {\bibfnamefont
  {R.}~\bibnamefont {Sabatini}}, \bibinfo {author} {\bibfnamefont
  {B.}~\bibnamefont {Santra}}, \bibinfo {author} {\bibfnamefont
  {M.}~\bibnamefont {Schlipf}}, \bibinfo {author} {\bibfnamefont {A.~P.}\
  \bibnamefont {Seitsonen}}, \bibinfo {author} {\bibfnamefont {A.}~\bibnamefont
  {Smogunov}}, \bibinfo {author} {\bibfnamefont {I.}~\bibnamefont {Timrov}},
  \bibinfo {author} {\bibfnamefont {T.}~\bibnamefont {Thonhauser}}, \bibinfo
  {author} {\bibfnamefont {P.}~\bibnamefont {Umari}}, \bibinfo {author}
  {\bibfnamefont {N.}~\bibnamefont {Vast}}, \bibinfo {author} {\bibfnamefont
  {X.}~\bibnamefont {Wu}}, \ and\ \bibinfo {author} {\bibfnamefont
  {S.}~\bibnamefont {Baroni}},\ }\href {\doibase 10.1088/1361-648X/aa8f79}
  {\bibfield  {journal} {\bibinfo  {journal} {Journal of Physics: Condensed
  Matter}\ }\textbf {\bibinfo {volume} {29}},\ \bibinfo {pages} {465901}
  (\bibinfo {year} {2017})}\BibitemShut {NoStop}%
\bibitem [{\citenamefont {Perdew}\ \emph {et~al.}(1996)\citenamefont {Perdew},
  \citenamefont {Burke},\ and\ \citenamefont
  {Ernzerhof}}]{perdew1996generalized}%
  \BibitemOpen
  \bibfield  {author} {\bibinfo {author} {\bibfnamefont {J.~P.}\ \bibnamefont
  {Perdew}}, \bibinfo {author} {\bibfnamefont {K.}~\bibnamefont {Burke}}, \
  and\ \bibinfo {author} {\bibfnamefont {M.}~\bibnamefont {Ernzerhof}},\
  }\href@noop {} {\bibfield  {journal} {\bibinfo  {journal} {Physical review
  letters}\ }\textbf {\bibinfo {volume} {77}},\ \bibinfo {pages} {3865}
  (\bibinfo {year} {1996})}\BibitemShut {NoStop}%
\bibitem [{\citenamefont {{van Setten}}\ \emph {et~al.}(2018)\citenamefont
  {{van Setten}}, \citenamefont {Giantomassi}, \citenamefont {Bousquet},
  \citenamefont {Verstraete}, \citenamefont {Hamann}, \citenamefont {Gonze},\
  and\ \citenamefont {Rignanese}}]{VANSETTEN201839}%
  \BibitemOpen
  \bibfield  {author} {\bibinfo {author} {\bibfnamefont {M.}~\bibnamefont {{van
  Setten}}}, \bibinfo {author} {\bibfnamefont {M.}~\bibnamefont {Giantomassi}},
  \bibinfo {author} {\bibfnamefont {E.}~\bibnamefont {Bousquet}}, \bibinfo
  {author} {\bibfnamefont {M.}~\bibnamefont {Verstraete}}, \bibinfo {author}
  {\bibfnamefont {D.}~\bibnamefont {Hamann}}, \bibinfo {author} {\bibfnamefont
  {X.}~\bibnamefont {Gonze}}, \ and\ \bibinfo {author} {\bibfnamefont {G.-M.}\
  \bibnamefont {Rignanese}},\ }\href {\doibase
  https://doi.org/10.1016/j.cpc.2018.01.012} {\bibfield  {journal} {\bibinfo
  {journal} {Computer Physics Communications}\ }\textbf {\bibinfo {volume}
  {226}},\ \bibinfo {pages} {39} (\bibinfo {year} {2018})}\BibitemShut
  {NoStop}%
\bibitem [{\citenamefont {Song}\ \emph {et~al.}(2013)\citenamefont {Song},
  \citenamefont {Giorgi}, \citenamefont {Yamashita},\ and\ \citenamefont
  {Hirao}}]{song2013communication}%
  \BibitemOpen
  \bibfield  {author} {\bibinfo {author} {\bibfnamefont {J.-W.}\ \bibnamefont
  {Song}}, \bibinfo {author} {\bibfnamefont {G.}~\bibnamefont {Giorgi}},
  \bibinfo {author} {\bibfnamefont {K.}~\bibnamefont {Yamashita}}, \ and\
  \bibinfo {author} {\bibfnamefont {K.}~\bibnamefont {Hirao}},\ }\href@noop {}
  {\bibfield  {journal} {\bibinfo  {journal} {The Journal of chemical physics}\
  }\textbf {\bibinfo {volume} {138}} (\bibinfo {year} {2013})}\BibitemShut
  {NoStop}%
\bibitem [{\citenamefont {Aryasetiawan}\ and\ \citenamefont
  {Gunnarsson}(1998)}]{aryasetiawan1998gw}%
  \BibitemOpen
  \bibfield  {author} {\bibinfo {author} {\bibfnamefont {F.}~\bibnamefont
  {Aryasetiawan}}\ and\ \bibinfo {author} {\bibfnamefont {O.}~\bibnamefont
  {Gunnarsson}},\ }\href@noop {} {\bibfield  {journal} {\bibinfo  {journal}
  {Rep. Prog. Phys.}\ }\textbf {\bibinfo {volume} {61}},\ \bibinfo {pages}
  {237} (\bibinfo {year} {1998})}\BibitemShut {NoStop}%
\bibitem [{\citenamefont {Godby}\ and\ \citenamefont {Needs}(1989)}]{PPA}%
  \BibitemOpen
  \bibfield  {author} {\bibinfo {author} {\bibfnamefont {R.~W.}\ \bibnamefont
  {Godby}}\ and\ \bibinfo {author} {\bibfnamefont {R.~J.}\ \bibnamefont
  {Needs}},\ }\href {\doibase 10.1103/PhysRevLett.62.1169} {\bibfield
  {journal} {\bibinfo  {journal} {Phys. Rev. Lett.}\ }\textbf {\bibinfo
  {volume} {62}},\ \bibinfo {pages} {1169} (\bibinfo {year}
  {1989})}\BibitemShut {NoStop}%
\bibitem [{\citenamefont {Gant}\ \emph {et~al.}(2022)\citenamefont {Gant},
  \citenamefont {Haber}, \citenamefont {Filip}, \citenamefont {Sagredo},
  \citenamefont {Wing}, \citenamefont {Ohad}, \citenamefont {Kronik},\ and\
  \citenamefont {Neaton}}]{Gant2022}%
  \BibitemOpen
  \bibfield  {author} {\bibinfo {author} {\bibfnamefont {S.~E.}\ \bibnamefont
  {Gant}}, \bibinfo {author} {\bibfnamefont {J.~B.}\ \bibnamefont {Haber}},
  \bibinfo {author} {\bibfnamefont {M.~R.}\ \bibnamefont {Filip}}, \bibinfo
  {author} {\bibfnamefont {F.}~\bibnamefont {Sagredo}}, \bibinfo {author}
  {\bibfnamefont {D.}~\bibnamefont {Wing}}, \bibinfo {author} {\bibfnamefont
  {G.}~\bibnamefont {Ohad}}, \bibinfo {author} {\bibfnamefont {L.}~\bibnamefont
  {Kronik}}, \ and\ \bibinfo {author} {\bibfnamefont {J.~B.}\ \bibnamefont
  {Neaton}},\ }\href {\doibase 10.1103/PhysRevMaterials.6.053802} {\bibfield
  {journal} {\bibinfo  {journal} {Phys. Rev. Mater.}\ }\textbf {\bibinfo
  {volume} {6}},\ \bibinfo {pages} {053802} (\bibinfo {year}
  {2022})}\BibitemShut {NoStop}%
\bibitem [{\citenamefont {Freysoldt}\ \emph {et~al.}(2014)\citenamefont
  {Freysoldt}, \citenamefont {Grabowski}, \citenamefont {Hickel}, \citenamefont
  {Neugebauer}, \citenamefont {Kresse}, \citenamefont {Janotti},\ and\
  \citenamefont {Van~de Walle}}]{rev_freysoldt_defects}%
  \BibitemOpen
  \bibfield  {author} {\bibinfo {author} {\bibfnamefont {C.}~\bibnamefont
  {Freysoldt}}, \bibinfo {author} {\bibfnamefont {B.}~\bibnamefont
  {Grabowski}}, \bibinfo {author} {\bibfnamefont {T.}~\bibnamefont {Hickel}},
  \bibinfo {author} {\bibfnamefont {J.}~\bibnamefont {Neugebauer}}, \bibinfo
  {author} {\bibfnamefont {G.}~\bibnamefont {Kresse}}, \bibinfo {author}
  {\bibfnamefont {A.}~\bibnamefont {Janotti}}, \ and\ \bibinfo {author}
  {\bibfnamefont {C.~G.}\ \bibnamefont {Van~de Walle}},\ }\href {\doibase
  10.1103/RevModPhys.86.253} {\bibfield  {journal} {\bibinfo  {journal} {Rev.
  Mod. Phys.}\ }\textbf {\bibinfo {volume} {86}},\ \bibinfo {pages} {253}
  (\bibinfo {year} {2014})}\BibitemShut {NoStop}%
\bibitem [{\citenamefont {Zhang}\ and\ \citenamefont
  {Northrup}(1991)}]{zhang_northrup_formation_en}%
  \BibitemOpen
  \bibfield  {author} {\bibinfo {author} {\bibfnamefont {S.~B.}\ \bibnamefont
  {Zhang}}\ and\ \bibinfo {author} {\bibfnamefont {J.~E.}\ \bibnamefont
  {Northrup}},\ }\href {\doibase 10.1103/PhysRevLett.67.2339} {\bibfield
  {journal} {\bibinfo  {journal} {Phys. Rev. Lett.}\ }\textbf {\bibinfo
  {volume} {67}},\ \bibinfo {pages} {2339} (\bibinfo {year}
  {1991})}\BibitemShut {NoStop}%
\bibitem [{\citenamefont {Freysoldt}\ \emph {et~al.}(2009)\citenamefont
  {Freysoldt}, \citenamefont {Neugebauer},\ and\ \citenamefont
  {Van\;de\;Walle}}]{freysoldt_phys_rev_lett}%
  \BibitemOpen
  \bibfield  {author} {\bibinfo {author} {\bibfnamefont {C.}~\bibnamefont
  {Freysoldt}}, \bibinfo {author} {\bibfnamefont {J.}~\bibnamefont
  {Neugebauer}}, \ and\ \bibinfo {author} {\bibfnamefont {C.~G.}\ \bibnamefont
  {Van\;de\;Walle}},\ }\href {\doibase 10.1103/PhysRevLett.102.016402}
  {\bibfield  {journal} {\bibinfo  {journal} {Phys. Rev. Lett.}\ }\textbf
  {\bibinfo {volume} {102}},\ \bibinfo {pages} {016402} (\bibinfo {year}
  {2009})}\BibitemShut {NoStop}%
\bibitem [{\citenamefont {Freysoldt}\ \emph {et~al.}(2011)\citenamefont
  {Freysoldt}, \citenamefont {Neugebauer},\ and\ \citenamefont
  {Van{\;}de{\;}Walle}}]{Freysoldt_phys_stat_b}%
  \BibitemOpen
  \bibfield  {author} {\bibinfo {author} {\bibfnamefont {C.}~\bibnamefont
  {Freysoldt}}, \bibinfo {author} {\bibfnamefont {J.}~\bibnamefont
  {Neugebauer}}, \ and\ \bibinfo {author} {\bibfnamefont {C.~G.}\ \bibnamefont
  {Van{\;}de{\;}Walle}},\ }\href {\doibase
  https://doi.org/10.1002/pssb.201046289} {\bibfield  {journal} {\bibinfo
  {journal} {physica status solidi (b)}\ }\textbf {\bibinfo {volume} {248}},\
  \bibinfo {pages} {1067} (\bibinfo {year} {2011})},\ \Eprint
  {http://arxiv.org/abs/https://onlinelibrary.wiley.com/doi/pdf/10.1002/pssb.201046289}
  {https://onlinelibrary.wiley.com/doi/pdf/10.1002/pssb.201046289} \BibitemShut
  {NoStop}%
\bibitem [{\citenamefont {Freysoldt}(2022)}]{sxdefectalign}%
  \BibitemOpen
  \bibfield  {author} {\bibinfo {author} {\bibfnamefont {C.}~\bibnamefont
  {Freysoldt}},\ }\href {https://sxrepo.mpie.de/projects/sphinx-add-ons/files}
  {\enquote {\bibinfo {title} {{Sphynx} add-ons},}\ } (\bibinfo {year}
  {2022})\BibitemShut {NoStop}%
\bibitem [{\citenamefont {Kumagai}\ and\ \citenamefont
  {Oba}(2014)}]{kumagai_oba_defects}%
  \BibitemOpen
  \bibfield  {author} {\bibinfo {author} {\bibfnamefont {Y.}~\bibnamefont
  {Kumagai}}\ and\ \bibinfo {author} {\bibfnamefont {F.}~\bibnamefont {Oba}},\
  }\href {\doibase 10.1103/PhysRevB.89.195205} {\bibfield  {journal} {\bibinfo
  {journal} {Phys. Rev. B}\ }\textbf {\bibinfo {volume} {89}},\ \bibinfo
  {pages} {195205} (\bibinfo {year} {2014})}\BibitemShut {NoStop}%
\bibitem [{\citenamefont {Shang}\ \emph {et~al.}(2007)\citenamefont {Shang},
  \citenamefont {Wang}, \citenamefont {Arroyave},\ and\ \citenamefont
  {Liu}}]{pure_boron_alpha_vs_beta}%
  \BibitemOpen
  \bibfield  {author} {\bibinfo {author} {\bibfnamefont {S.}~\bibnamefont
  {Shang}}, \bibinfo {author} {\bibfnamefont {Y.}~\bibnamefont {Wang}},
  \bibinfo {author} {\bibfnamefont {R.}~\bibnamefont {Arroyave}}, \ and\
  \bibinfo {author} {\bibfnamefont {Z.-K.}\ \bibnamefont {Liu}},\ }\href
  {\doibase 10.1103/PhysRevB.75.092101} {\bibfield  {journal} {\bibinfo
  {journal} {Phys. Rev. B}\ }\textbf {\bibinfo {volume} {75}},\ \bibinfo
  {pages} {092101} (\bibinfo {year} {2007})}\BibitemShut {NoStop}%
\bibitem [{\citenamefont {Abdi}\ \emph {et~al.}(2018)\citenamefont {Abdi},
  \citenamefont {Chou}, \citenamefont {Gali},\ and\ \citenamefont
  {Plenio}}]{Abdi2018}%
  \BibitemOpen
  \bibfield  {author} {\bibinfo {author} {\bibfnamefont {M.}~\bibnamefont
  {Abdi}}, \bibinfo {author} {\bibfnamefont {J.-P.}\ \bibnamefont {Chou}},
  \bibinfo {author} {\bibfnamefont {A.}~\bibnamefont {Gali}}, \ and\ \bibinfo
  {author} {\bibfnamefont {M.~B.}\ \bibnamefont {Plenio}},\ }\href {\doibase
  10.1021/acsphotonics.7b01442} {\bibfield  {journal} {\bibinfo  {journal} {ACS
  Photonics}\ }\textbf {\bibinfo {volume} {5}},\ \bibinfo {pages} {1967}
  (\bibinfo {year} {2018})},\ \Eprint
  {http://arxiv.org/abs/https://doi.org/10.1021/acsphotonics.7b01442}
  {https://doi.org/10.1021/acsphotonics.7b01442} \BibitemShut {NoStop}%
\bibitem [{\citenamefont {Janotti}\ and\ \citenamefont {Van~de
  Walle}(2007)}]{janotti_van_de_walle_zno}%
  \BibitemOpen
  \bibfield  {author} {\bibinfo {author} {\bibfnamefont {A.}~\bibnamefont
  {Janotti}}\ and\ \bibinfo {author} {\bibfnamefont {C.~G.}\ \bibnamefont
  {Van~de Walle}},\ }\href {\doibase 10.1103/PhysRevB.76.165202} {\bibfield
  {journal} {\bibinfo  {journal} {Phys. Rev. B}\ }\textbf {\bibinfo {volume}
  {76}},\ \bibinfo {pages} {165202} (\bibinfo {year} {2007})}\BibitemShut
  {NoStop}%
\bibitem [{\citenamefont {Laaksonen}\ \emph {et~al.}(2008)\citenamefont
  {Laaksonen}, \citenamefont {Ganchenkova},\ and\ \citenamefont
  {Nieminen}}]{Laaksonen_2009}%
  \BibitemOpen
  \bibfield  {author} {\bibinfo {author} {\bibfnamefont {K.}~\bibnamefont
  {Laaksonen}}, \bibinfo {author} {\bibfnamefont {M.~G.}\ \bibnamefont
  {Ganchenkova}}, \ and\ \bibinfo {author} {\bibfnamefont {R.~M.}\ \bibnamefont
  {Nieminen}},\ }\href {\doibase 10.1088/0953-8984/21/1/015803} {\bibfield
  {journal} {\bibinfo  {journal} {Journal of Physics: Condensed Matter}\
  }\textbf {\bibinfo {volume} {21}},\ \bibinfo {pages} {015803} (\bibinfo
  {year} {2008})}\BibitemShut {NoStop}%
\bibitem [{\citenamefont {Gali}\ \emph {et~al.}(2008)\citenamefont {Gali},
  \citenamefont {Fyta},\ and\ \citenamefont {Kaxiras}}]{nv_diamon_struct}%
  \BibitemOpen
  \bibfield  {author} {\bibinfo {author} {\bibfnamefont {A.}~\bibnamefont
  {Gali}}, \bibinfo {author} {\bibfnamefont {M.}~\bibnamefont {Fyta}}, \ and\
  \bibinfo {author} {\bibfnamefont {E.}~\bibnamefont {Kaxiras}},\ }\href
  {\doibase 10.1103/PhysRevB.77.155206} {\bibfield  {journal} {\bibinfo
  {journal} {Phys. Rev. B}\ }\textbf {\bibinfo {volume} {77}},\ \bibinfo
  {pages} {155206} (\bibinfo {year} {2008})}\BibitemShut {NoStop}%
\bibitem [{\citenamefont {Ádám Gali}(2019)}]{Gali_review}%
  \BibitemOpen
  \bibfield  {author} {\bibinfo {author} {\bibnamefont {Ádám Gali}},\ }\href
  {\doibase doi:10.1515/nanoph-2019-0154} {\bibfield  {journal} {\bibinfo
  {journal} {Nanophotonics}\ }\textbf {\bibinfo {volume} {8}},\ \bibinfo
  {pages} {1907} (\bibinfo {year} {2019})}\BibitemShut {NoStop}%
\end{thebibliography}%

\end{document}